\documentclass{aa} 
\usepackage{natbib}
\bibpunct{(}{)}{;}{a}{}{,} 
\usepackage{graphicx}
\usepackage{multirow}
\usepackage{txfonts}
\newcommand{\msun}{{\rm M_\odot}}
\newcommand{\tco}{\ifmmode {^{13}{\rm CO}} \else {$^{13}{\rm CO}$}\fi}
\newcommand{\dco}{\ifmmode {^{12}{\rm CO}} \else {$^{12}{\rm CO}$}\fi}
\newcommand{\cdo}{\ifmmode {{\rm C}^{18}{\rm O}} \else {${\rm C}^{18}{\rm O}$}\fi}
\newcommand{\htco}{\ifmmode {{\rm H}^{13}{\rm CO}^{+} } \else {${\rm H}^{13}
{\rm CO}^{+}$ }\fi}
\newcommand{\hco}{\ifmmode {{\rm H}^{12}{\rm CO}^{+} } \else {${\rm H}^{12}
{\rm CO}^{+}$ }\fi}
\newcommand{\juz}{\ifmmode {{\rm J}=1\rightarrow 0} \else
{J=1$\rightarrow$0}\fi}
\newcommand{\jdu}{\ifmmode {{\rm J}=2\rightarrow 1} \else
{J=2$\rightarrow$1}\fi}
\newcommand{\jtd}{\ifmmode {{\rm J}=3\rightarrow 2} \else
{J=3$\rightarrow$2} \fi}
\newcommand{\jcq}{\ifmmode {{\rm J}=5\!\rightarrow\!4} \else
{${\rm J}=5\!\rightarrow\!4$} \fi}
\newcommand{\dcod}{\ifmmode {^{12}{\rm CO_{disk}}} \else {$^{12}{\rm CO_{disk}}$}\fi}
\newcommand{\dcos}{\ifmmode {^{12}{\rm CO_{spiral}}} \else {$^{12}{\rm CO_{spiral}}$}\fi}

\begin{document}
  \title{The circumstellar disk of AB Aurigae: evidence for envelope accretion at late stages of star formation?\thanks{Based on observations carried out with the IRAM 30-m telescope and Plateau de Bure
interferometer.  IRAM is supported by INSU/CNRS (France), MPG (Germany), and IGN (Spain).} \thanks{Based on observations carried out with the Submillimeter Array (SMA). The SMA is a joint project between the Smithsonian Astrophysical Observatory and the Academia Sinica Institute of Astronomy and Astrophysics and is funded by the Smithsonian Institution and the Academia Sinica.}}

   \subtitle{ }

   \author{Ya-Wen~Tang \inst{1,2,3}  \and St\'{e}phane Guilloteau \inst{1,2} \and Vincent~Pi\'etu \inst{4}  \and Anne~Dutrey \inst{1,2}
   \and Nagayoshi~Ohashi \inst{3,5} \and Paul T.P.~Ho \inst{3,6}}

   \institute{   Universit\'e de Bordeaux, Observatoire Aquitain des Sciences de l'Univers,
2 rue de l'Observatoire, BP 89, F-33271 Floirac Cedex, France \and
CNRS, UMR 5804, Laboratoire d'Astrophysique de Bordeaux,
2 rue de l'Observatoire, BP 89, F-33271 Floirac Cedex, France
\and
   Institute of Astronomy and Astrophysics, Academia Sinica,
Taiwan, R.O.C
\and
IRAM, 300 rue de la piscine, F-38406 Saint Martin d'H\`eres Cedex, France
\and 
Subaru Telescope, 650 North A'ohoku Place, Hilo, HI 96720, USA  
\and
Harvard Smithsonian Center for Astrophysics, 60 Garden Street, Cambridge, MA 02138, USA
 }
   \date{Received xxx ; accepted xxx}


  \abstract
  {}
  {The circumstellar disk of AB Aurigae has garnered strong attention owing to the apparent existence of spirals at a relatively young stage and also the asymmetric disk traced in thermal dust emission. However, the physical conditions of the spirals are still not well understood. The origin of the asymmetric thermal emission is unclear.
  }
  {We observed the disk at 230 GHz (1.3 mm) in both the continuum and the spectral line \dco~\jdu~with IRAM 30-m, the Plateau de Bure interferometer, and the Submillimeter Array to sample all spatial scales from 0$\farcs$37 to about 50$''$. To combine the data obtained from these telescopes, several methods and calibration issues were checked and discussed.}
  {The 1.3 mm continuum (dust) emission is resolved into inner disk and outer ring. The emission from the dust ring is highly asymmetric in azimuth, with intensity variations by a factor 3. Molecular gas at high velocities traced by the CO line is detected next to the stellar location. The inclination angle of the disk is found to decrease toward the center. On a larger scale, based on the intensity weighted dispersion and the integrated intensity map of \dco~\jdu, four spirals are identified, where two of them are also detected in the near infrared.  The total gas mass of the 4 spirals ($M_{\rm spiral}$) 
  is 10$^{-7}$ $<$ $M_{\rm spiral}$ $<$ 10$^{-5}$ $\msun$, which is 3 orders of magnitude smaller than the mass of the gas ring. Surprisingly, the CO gas inside the spiral is apparently counter-rotating with respect to the CO disk, and it only exhibits small radial motion.}
  {The wide gap, the warped disk, and the asymmetric dust ring suggest that there is an undetected companion with a mass of 0.03 $\msun$ at a radius of 45 AU. 
The different spirals would, however, require multiple perturbing bodies. While viable from an energetic point of view,
  this mechanism cannot explain the apparent counter-rotation of the gas in the spirals. Although an hypothetical fly-by cannot be ruled out, the most likely explanation of the AB Aurigae system may be inhomogeneous accretion well above or below the main disk plane from the remnant envelope, which can explain both the rotation and large-scale motions detected  with the 30-m image.}


   \maketitle


\section{Introduction}
Circumstellar disks are commonly found around pre-main-sequence stars \citep{Strom+etal_1989,Beckwith+etal_1990}.
They are believed to play an important role in processing the excess angular momentum of cloud cores and in allowing the accretion to proceed inward toward the central star radially via gravitational instabilities \citep[see the review by][]{Durisen+etal_2007} or magnetorotational instability \citep{Balbus1991,Balbus1998}, or vertically \citep[and references therein]{Konigl2011}.
Within these disks, planetary systems are expected to form in the early \citep{Inutsuka+etal_2010} or later evolutionary stages.
As a result, the structure and kinematics of the disks of pre-main-sequence stars provide  clues to the linkage between star and planet formation.

With the improvement in angular resolution and in sensitivity, CO maps reveal that these disks usually exhibit Keplerian kinematics
on scales of 100s AU \citep[e.g.][]{Pietu+etal_2007}. In the meantime, more and more circumstellar disks are found to have
complex structures, such as the large cavities found in HD 135344B \citep{Brown+etal_2009}, MWC\,758 \citep{Isella+etal_2010}, LkCa 15 \citep{Pietu+etal_2006},
and HH\,30 \citep{Guilloteau+etal_2008}. A more thorough analysis of the structure and kinematics of circumstellar disks are therefore important for providing constraints on the physical conditions of planetary system formation.

Located at about 140~pc, AB Aurigae (hereafter AB Aur) is one of the closer
Herbig Ae stars of spectral type of  A0-A1 \citep{Hernandez+etal_2004}.
Modeling of its mid-infrared spectral energy distribution (SED)
by \citet{Meeus+etal_2001} and \citet{Bouwman+etal_2000} shows that AB Aur belongs to
Group I, where the SED can be reconstructed by the combination of a power law and a black body, so this young star is usually considered as
the prototype of the Herbig Ae star surrounded by a large
flaring disk.

Indeed, AB Aur is surrounded by a large amount of material that
is observed from the near infrared \citep[NIR; e.g. HST-STIS imaging by][]{Grady+etal_1999}
up to the mm domain \citep{Pietu+etal_2005,Corder+etal_2005}.
The NIR images have revealed a flattened reflection nebula,
close to pole-on, which extends up to $r \sim 1300$\,AU from
the star. \citet{Fukagawa+etal_2004} also made deep NIR images of AB Aur using
the Coronographic Imager Adaptive Optics systems on the Subaru
telescope. Their observations revealed spiral
features within the circumstellar matter.
Interestingly, part of these
spirals have also been marginally detected in CO 3-2 line emission by \citet{Lin+etal_2006} using the Submillimeter Array (SMA).
Finally, independent studies of the CO line kinematics have revealed
that the outer disk structure is
{\it in sub-Keplerian rotation} \citep{Pietu+etal_2005,Lin+etal_2006}, unlike what has been found in other similar systems \citep[e.g.][]{Simon+etal_2000}.
Since the disk does not appear massive enough to be self-gravitating,
one plausible explanation would be the youth of the system.
The Plateau de Bure interferometer (PdBI) images from \citet{Pietu+etal_2005} at 1.3mm and in CO 2-1 also show that
 the material is truncated toward the center with a ring-like distribution with an inner radius
 around $70-100$\,AU. The origin of such a cavity has not yet been identified.
 This may be due to a substellar or a very low-mass stellar companion located at $\sim 20-40$~AU
 from the primary, as suggested by the recent very large array (VLA) observations at the wavelengths of 3.6\,cm by \citet{Rodriguez+etal_2007}.

Therefore, both the gas and dust orbiting around AB Aur present several puzzling features,
which include spirals, (sub)Keplerian motions, and an inner cavity.
This object appears very different from the simple
picture of a flaring, Keplerian gas and dust disk. It may
represent the real complexity of an evolving young system in which
planet formation has already started.
In this paper we attempt to improve our physical understanding of AB Aur by conducting a joint IRAM-SMA project in order to better define the mm/submm properties of the source. Sections 2 and 3 present the observations and results. We discuss the implications of
the observed properties in Sect. 4 and summarize in Sect. 5.

\section{Observations and data analysis}
We imaged the 1.3 mm continuum emission by utilizing the highest angular resolution (0$\farcs$3) data from the PdBI, together with the short-spacing information from the SMA. The gas kinematics were studied with \dco~\jdu~emission using the IRAM 30-m, PdBI, and SMA.

\subsection{IRAM 30-m data}

The IRAM 30-m \dco~\jdu~data were obtained on Sep 9-10, 2009. The receivers were tuned in single sideband, with a
measured sideband rejection of about 13 dB. We used a spectral resolution of 40 kHz (0.052 km s$^{-1}$) for the spectra.
We used the on-the-fly mapping technique. A fixed position near AB Aur was used as the reference position to subtract the background. This position was, however, not free of CO emission: it was observed in frequency switching to recover
the full flux, and the spectrum was added back to the data obtained in the on-the-fly processing. Consequently, the 30-m is not
missing any spectral emission. However, the continuum level is lost due to the linear baseline, which was fitted to remove residual
atmospheric contamination.

Pointing calibration was done in two steps. First, pointing was checked every two hours at the telescope.
Focus was also monitored and found to be stable. Pointing accuracy was then refined by obtaining a series of independent images, each taking about half an hour of elapsed time. Each image
was then recentered (through cross-correlation) on the nominal position determined from the PdBI data,
using the high-velocity channels that only come from the disk emission (see later discussion).
This allows us to correct the 30-m pointing errors as a function of time.
Finally, each spectrum was corrected for the determined pointing offsets (using linear interpolation in time
from the values derived for each map), and the final images were produced.

The overall calibration procedure guarantees a registration better than 1$''$ between the
30-m and interferometer data, and an amplitude calibration precision better than 10 \%.


\begin{table*}
\caption{Observation parameters}
\centering
\begin{tabular}{ll | l | l l |l l}
\hline
\multicolumn{2}{c}{\multirow{2}{*}{Parameters}} & \multicolumn{1}{c}{\multirow{2}{*}{Obs. scheme}} & \multicolumn{2}{c}{RMS} & \multicolumn{2}{c}{Angular resolution} \\
     & & & (mJy/beam) & (K) & size (\arcsec $\times$ \arcsec) & PA (\degr) \\ \hline\hline
 30-m  & CO & on-the-fly & 1150  & 0.2  & 12$\times$12 & - \\\hline
\multirow{2}{*}{PdBI} & 1.3 mm & seven-field & 0.23 & 0.027   & 0.52$\times$0.37 & 24 \\
                      & CO     & mosaic  & 22   & 2.2  & 0.56$\times$0.42 & 27 \\ \hline
\multirow{2}{*}{SMA}  & 1.3 mm & single & 1 & 8$\times$10$^{-4}$ & 7.4$\times$4.4 & 166 \\
                      & CO     & pointing & 110 & 0.09 & 7.4$\times$4.4 & 166 \\\hline
\end{tabular}
\label{tab:obs}
\end{table*}

\subsection{Plateau de Bure interferometer mosaic}

The high angular resolution data were obtained in mosaic centered on AB Aur, observed with long baselines in the AB configuration, on Jan 30, Feb 20, and Mar 14, 2009, and on Jan 22 and Feb 2, 2010, in both continuum and in \dco~\jdu. The seven-field mosaic uses a hexagonal pattern with
field separated by 11$\arcsec$, about half of the half-power primary beam of the 15-m dishes at this frequency (23$\arcsec$).
The baselines range from 45 m to 760 m. These newly obtained data were supplemented by the existing compact configuration (CD) data from \citet{Pietu+etal_2005}, which cover one field (central field) and baseline lengths from 15 to 174 m.

In all cases, the correlator was set with two units covering the \dco~\jdu~line at high spectral resolution (80 kHz, or 0.104 km s$^{-1}$).
The remaining six units were used to observe the continuum emission covering a bandwidth of 0.96 GHz in HH and VV polarization.
All data were processed using the CLIC software in the GILDAS package.
Phase and amplitude calibration were performed using the quasars 3C111 and 0518+134, while the flux calibration used MWC\,349 as a reference.

The combined data set results in an angular resolution of $0\farcs52 \times 0\farcs37$ at positional angle (PA) of $24^\circ$ and $0\farcs56 \times 0\farcs42$ at PA of $27^\circ$ for the 1.3 mm continuum and \dco~\jdu, respectively.
The noise level of the 1.3 mm continuum, $\sigma_{\rm I}$, is 0.23 mJy beam$^{-1}$. The noise level for the \dco~\jdu~line is 22 mJy beam$^{-1}$ (2.2 K) at the spectral resolution of 0.203 km s$^{-1}$, which is the resolution of the maps shown later in this paper.

\subsection{SMA observations and data reduction}

The PdBI mosaic lacks short spacings below 45 m except for the central field,
and the IRAM 30-m can only provide reliable spacings up to 15 m. Furthermore, the 30-m data do not give any continuum information.
We therefore observed in \dco~ \jdu~ and in continuum at 1.3\,mm, a single field with the SMA in the subcompact configuration, to sample the deprojected baseline range 6 to 68 m (4.4  to 52 k$\lambda$).
The correlator was set to cover the \dco~\jdu~line in the upper sideband with high spectral resolution (50.8 kHz, or 0.066 km s$^{-1}$).
The rest of the chunks with the equivalent bandwidth of 3.28 GHz were set for the continuum.

The short spacing data from the SMA were obtained on Sep 14, 2009 under weather conditions of $\tau$ = 0.1 at 225 GHz and with humidity between
7\% and 20\%. The data were calibrated and reduced using the {\it MIR}\ software. {\it MIR} is a software package to reduce data taken with the SMA. The bandpass, flux, and gain calibrator are 3C454.3, Uranus, and
3C111, respectively. The phase scatter on 3C111 is $\pm$ 20$\degr$.
The angular resolution of the SMA dataset is 7$\farcs$4 $\times$ 4$\farcs$4 at PA of 166$\degr$, and the noise level is 1 mJy beam$^{-1}$.
The half-power field of view of the SMA at this frequency is $55\arcsec$, which is sufficient to cover the field covered with the PdBI with a seven-field mosaic.

\subsection{Data merging}

All spectral data were resampled to a 0.203 km s$^{-1}$ spectral resolution.
Corrections for the proper motion of AB Aur,  (1.7, -24.0) mas yr$^{-1}$ from Hipparcos, were applied to all data, and all positions referred to equinox J2000. From the existing observations, several combined datasets were produced. In this paper, we present and discuss the following cases.

\begin{enumerate}

\item  \noindent The combined 30-m + SMA $uv$ data: they were imaged in a standard way. The resulting data cube has a
typical rms noise of 0.13 Jy beam$^{-1}$ in nearly empty channels. In the channels with strong emission, the rms
rises up to 0.39 Jy beam$^{-1}$, as a result of the effective dynamic range limitations due to phase (SMA data) or
pointing (30-m) errors, as well as uncertainties in intensity calibrations. 
The peak signal-to-noise ratio is about 160.

\item \noindent The PdBI mosaic: seven-field mosaic observed with medium, long, and very long baselines (up to 760 m) but lacking baselines shorter than 45 m. The central field has a much better $uv$ coverage because of the existing PdBI data from \citet{Pietu+etal_2005}.

\item \noindent The PdB+SMA+30-m data: the two previous data sets were combined using a specially designed
procedure. To do so, we utilized the fact that the SMA+30-m deconvolved image has high signal-to-noise.
We thus used it as a model for the sky emission to compute model visibilities and added these to the PdBI
$uv$ table for each field.  In this process, the respective primary beams of the SMA (centered) and PdBI
(different for each field) must be taken into account. In the combination, the choice of the $uv$ coverage
on which visibilities must be computed is rather arbitrary. At the minimum, we can use the sampling
defined by the SMA tracks completed with the 30-m spacings up to about 10 m. 
For the spacings beyond 10 m the SMA becomes more precise than the 30-m owing to the limited pointing accuracy of the latter. Alternatively, we can use any arbitrary sampling up to about 40 m, beyond which the SMA+30-m is no longer sensitive enough compared to the PdBI data.
The weights must also be adjusted to match the PdBI weights and to obtain a final
distribution of weights that is as uniform as possible; this reweighting necessarily implies some loss
in sensitivity. This overall process is strictly equivalent to considering that
the SMA+30-m deconvolved image (corrected for the primary beam) is like a 30-m image, except for the
different angular resolution. We experimented with several weightings and $uv$ samplings, and as expected,
the results are relatively insensitive to the (arbitrary) choices, provided we do not select
$uv$ spacings that are too extreme from each telescope.

\item \noindent
In case the signal-to-noise ratio in the SMA+30-m image is not high enough, a
different procedure may be used. Rather than retaining the complete image, we can instead use the Clean
components, as only noise should be left out. Each Clean component must be corrected for the different
primary beam attenuations (including the pointing center offsets for the PdBI). A $uv$ table can then
be created from these Clean components. We have, however, enough sensitivity in all cases that
this procedure gives results that are totally compatible with the one mentioned in item (3) above.
\end{enumerate}

Relative calibration of the data from these three telescopes was checked by comparing
visibilities from the CO channels (the continuum flux being too weak for this purpose).
This process did not reveal any major discrepancy, but had limited accuracy (20 \% level
at best) because of the complex spatial structures. Comparison of strictly redundant spacings
does not provide sufficient sensitivity. We thus kept the original flux calibration scale
of all data sets, which, given the observing techniques, should be more accurate than 10 \%.
In practice, our conclusions are fairly insensitive to the relative calibration errors, since the only combined data being used is for the continuum (see Sect.\ref{sec:sub:cont} to \ref{sec:sub:co}).

The maps and spectra of \dco~\jdu~ line presented in this paper are from PdBI data, unless it is explicitly mentioned.
All relative positions cited in this paper are with respect to the coordinate ($\alpha$, $\delta$)$_{\rm J2000}$ = (04$^h$55$^m$45$\farcs$84, +30$\degr$33$\arcmin$04$\farcs$30). Table 1 summarizes the observational parameters.

\section{Results and modeling}

\begin{figure*}[!ht]
\includegraphics[scale=0.6]{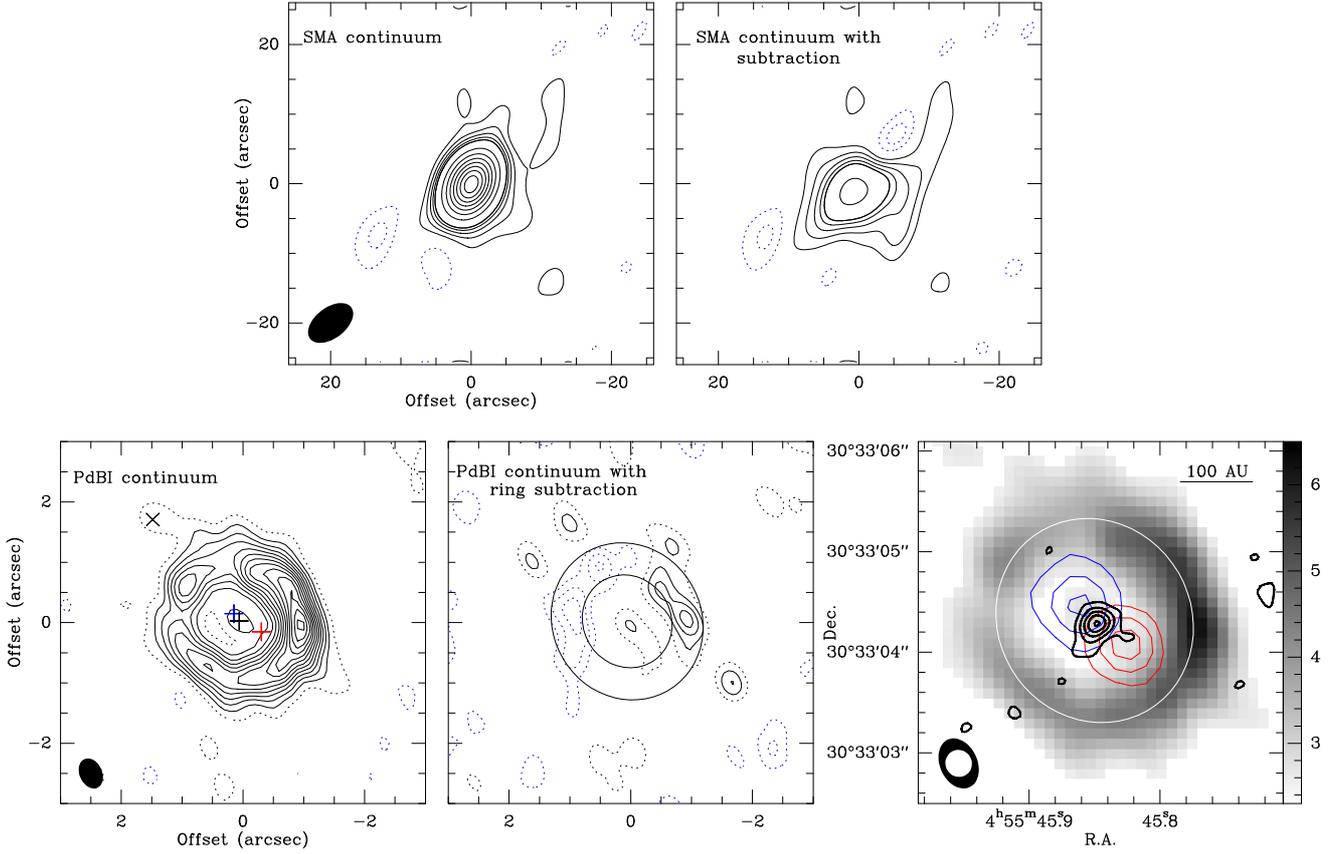}
\caption{Continuum emission maps at 1.3 mm of the AB Aur region. Top panels: Larger scale image obtained from the SMA data (left panel) and the residual image after removing the emission detected by the PdBI (right panel). Contour levels
are -2, 2, 4, 6, 8, 10 to 80 by 10 $\times$ 1 mJy per 7$\farcs$4$\times$4$\farcs$4 beam. Bottom panels: Higher angular resolution images obtained with the PdBI. Left panel: The blue, red, and black pluses mark the peaks of CO 2-1 highest blue-shifted, red-shifted peak, and the stellar location, respectively. 
The black cross marks the northeast peak of the 1.3 mm continuum. The dotted contours mark
the intensity of 2 $\sigma_{\rm I}$, where $\sigma_{\rm I}$= 0.23 mJy per 0$\farcs$52$\times$0$\farcs$37 beam. Black contours start from
4 $\sigma_{\rm I}$ in steps of 2 $\sigma_{\rm I}$.
Middle panel: Residual emission from the PdBI after removal of a best-fit smooth ring. The inner and outer radii are indicated. Contour levels are multiples of 2 $\sigma_{\rm I}$.
Right panel: the solid white ellipse marks the 1.3 mm peak ring with
a radius of 145 AU, $i$ of 23$\degr$ and PA of -30$\degr$ centered on the 1.3 mm stellar peak. The blue and red contours mark the CO 2-1 emission starting from and stepping in 66 mJy per 0$\farcs$56$\times$0$\farcs$42 beam at the velocity channels of 2 and 9.3 km s$^{-1}$, respectively.
The 3.6 cm continuum emission are in thick black contours at the level 3, 6, 9, 12, 15 $\times$ 9.3 $\mu$Jy per 0$\farcs$27$\times$0$\farcs$26 beam. The synthesized beam of 1.3 mm and 3.6 cm are labeled as black and white ellipses in the lower left corner, respectively.
}
\label{fig:continuum}
\end{figure*}

\subsection{Continuum observations}

Figure \ref{fig:continuum} is a montage of our 1.3 mm continuum maps obtained from the SMA and PdBI data.

\subsubsection{Large-scale continuum}
\label{sec:sub:cont}

For the SMA data (upper-left panel in Fig. \ref{fig:continuum}), the total flux is about 110 mJy and the rms noise 1 mJy beam$^{-1}$.
This value is only marginally lower than the single-dish measurement of 130 mJy, made at slightly higher frequency \citep{Henning+etal_1998}.
Although the emission appears localized toward AB Aur, it does not only originate in the circumstellar disk of AB Aur. The PdBI
data only recovers about 80 mJy of the total flux (see below). After removing the ring emission recovered by the PdBI from the SMA data, there is residual emission toward the south of the disk (upper-right panel of Fig. \ref{fig:continuum}).

The excess emission peaks about 1$\farcs$4 southeast of AB Aur, and half of the missing flux is within one SMA beam.
There is no significant structure on scales smaller than about $3''$. Such structures would have been recovered
by the PdBI, because it provides sufficient $uv$ coverage and sensitivity to $2-3''$ scales.
This excess emission is thus smooth on a scale of $3''$ or more. It presumably traces the remnant envelope of AB Aur.

With a dust absorption coefficient, $\rm \kappa$ = 0.02 cm$^2$ g$^{-1}$ (gas+dust) at 230 GHz, which is characteristic of disks \citep{Beckwith+etal_1990},
and a dust temperature of 30~K, the total gas mass obtained with the SMA would be $10^{-2}$ M$_{\sun}$. For more pristine, ISM-like dust,
the mass would be about four times higher. For the residual emission, the mean $N_{\rm H}$ is $1.3\times10^{23}$ (cm$^{-2}$).
Assuming the residual emission is spherically symmetric, the mean number volume density $n_{\rm H}$ is $5\times10^{6}$ (cm$^{-3}$).

\subsubsection{High angular resolution data}\label{sec:high angular resol}

At high angular resolution, the 1.3 mm continuum emission is resolved into three structures (bottom panels in Fig. \ref{fig:continuum}):
\begin{itemize}
\item a continuum peak at the central stellar position,
\item a continuum gap between the central peak and a dust ring,
\item a dust ring/ellipse.
\end{itemize}

The continuum peak near the center is detected at the level of 5.7 $\sigma_{\rm I}$ with a total flux of $1.3$~mJy.
It is not resolved within the 0$\farcs$37 (50 AU) interferometric beam. The mm continuum peak also
coincides (within the astrometric accuracy of our measurements,
about 0$\farcs$05) with the stellar position and the 3.6 cm emission peak (Fig. \ref{fig:continuum}, lower-right panel).
The flux density of the 3.6 cm emission is 0.2 mJy \citep{Rodriguez+etal_2007}. With a spectral
index of 0.6, characteristic of optically thick emission in an expanding ionized jet or wind \citep{Panagia+Felli_1975}, this would yield 1.5 mJy at 1.3 mm, so that the 1.3 mm signal from the star position could be purely due to an ionized
jet. On the other hand, optically thin free-free emission would have an essentially flat spectrum, so the
remaining $\approx 1.1$ mJy flux could come from dust emission from an inner disk.
A very small (2-3 AU), optically thick, and dust disk would provide enough flux.
A (gas+dust) mass of  4$\times$10$^{-5} \msun$ is required for such inner disk, assuming the temperature of 80 K and $\rm \kappa$ = 0.02 cm$^2$ g$^{-1}$. 
This mass is likely a lower limit since the dust emission should be partially optically thick.

The continuum gap (cavity) between this central peak and the dust ring has a width of
$\sim$0$\farcs$6, corresponding to about 90 AU. It appears more pronounced in the east
and southeast, but this is not statistically significant.

The dust ring/ellipse confirms the structure found by \citet{Pietu+etal_2005}.
With higher spatial resolution, we are able to further separate the emission from the central star.
For an assumed dust temperature $T(r) = 50 (r/100 \mathrm{AU})^{0.4}$ K (see the following paragraph)
and $\rm \kappa$ = 0.02 cm$^2$ g$^{-1}$, the ring mass is $5.0 \pm 0.5 \times 10^{-3} \msun$. With the same assumptions as in Sect. 3.1.1, the peak $N_{\rm H}$ and $n_{\rm H}$ of the dust ring are 7.3$\times$10$^{25}$ (cm$^{-2}$) and 7.0$\times$10$^{10}$ (cm$^{-3}$), respectively.
The intensity contrast on the dust ring is $\le$ 3.

\begin{figure}[!ht]
\includegraphics[width=8.0cm]{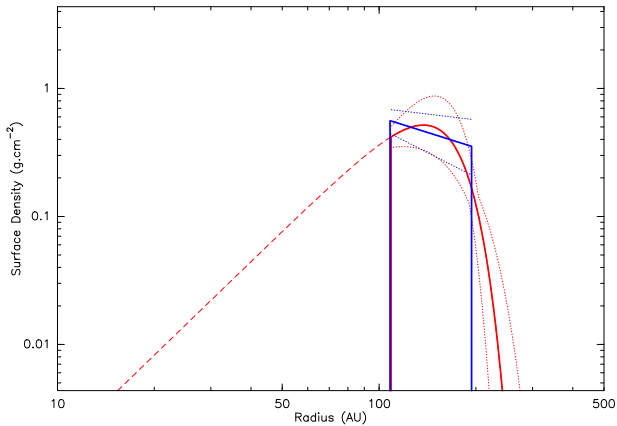}
\caption{Best-fit surface density distributions for the power-law (blue) and tapered-edge (red) models. Dotted curves represent the $\pm$ 1$\sigma$ of the best-fit value.
Without an inner edge, the tapered edge solution would provide too much flux from the inner
100 AU, as shown by the dashed line.}
\label{fig:dust_profile}
\end{figure}

The continuum emission is further fitted using DiskFit with an inclined, sharp-edged ring, as done for GG\,Tau \citep{Guilloteau+etal_1999}.
We fitted two different functionals: a truncated power law given by
\begin{equation}
  \Sigma (r) = \Sigma_0 (r/R_0)^{-p}   \mathrm{~~~~for~~~~} R_\mathrm{in} < r < R_\mathrm{out},
\end{equation}
and an exponentially tapered distribution given by
\begin{equation}
  \Sigma (r) = \Sigma_0 (r/R_0)^{-p} \exp (-(r/R_c)^{2-p})   \mathrm{~~~~for~~~~} R_\mathrm{in} < r.
\end{equation}
The best-fit of the dust ring parameters are listed in Table \ref{tab:cont}. 
Both functionals give similar results, in particular for the inner radius. The derived surface density profiles are given in Fig. \ref{fig:dust_profile}. The tapered edge
solution deserves specific comment. With the large inner radius also found in our best-fit solution, the high negative value found for the exponent is required to fit the steep decrease in surface brightness in the outer part.
A solution in which both the inner and outer parts are fit by a profile with no sharp inner radius is possible, but would
imply even more negative exponent values, below -4.5, a situation very similar to that found by \citet{Isella+etal_2010} for MWC 758. Besides being somewhat better than this solution, our fit demonstrates that the outer edge is rather
sharp and that the dust emission does not extend beyond about 250 AU.

%

In addition to these three structures, there is a northeast continuum peak 2$\farcs$2 away from the stellar position, corresponding to the ET peak identified by \citet{Lin+etal_2006} in their 0.88 mm continuum image.
It is in between CO S1 and S2 arms (see Sect. \ref{result:spiral}).
The intensity of this northeast peak is 0.76 mJy beam$^{-1}$ (3.3 $\sigma_{\rm I}$).
Because it is statistically not significant, we do not discuss this component further.

The center of the dust ring is at ($\alpha$,$\delta$)$_{\rm J2000}$=(4$^{h}$55$^{m}$45.85$^{s}$, 30${\degr}$33${\arcmin}$4$\farcs$32), consistent with the stellar location. The position angle (PA) of the rotation axis with respect to north of the best-fit ring is $\sim$-30$\degr$.
The inclination angle, $i$, is $\sim$23$\degr$ based on the ratio of minor axis to major axis assuming the ellipse is thin.
After subtracting the ring, the residual
image given in Fig. \ref{fig:continuum} (bottom-middle panel) shows a strong northsouth residual on the western side and a strong negative residual on the eastern side, suggesting that the emission on the dust ring is not azimuthally symmetric.

%

\begin{table}
\caption{Fitting parameters of dust ring}
\centering
\begin{tabular}{l|ll}
& Power law & Exponential tapering\\ \hline
Inner radius (AU) & $108 \pm 6$ & $109 \pm 4$ \\
Outer radius (AU) & $193 \pm 4$ & [ 350 ] \\
Critical radius (AU)  & [ $\infty$ ] & $157 \pm 4$ \\
$p$  & $0.8 \pm 0.5$ &  $-2.4 \pm 1.0$ \\
PA ($^\circ$) & $-31 \pm 6$ & $-27 \pm 9$ \\
$i$ ($^\circ$) & $20 \pm 3$ & $22 \pm 4$
\end{tabular}
\label{tab:cont}
\end{table}
%

%

\begin{figure*}[!ht]
	\centering
		\includegraphics[width=0.8\textwidth]{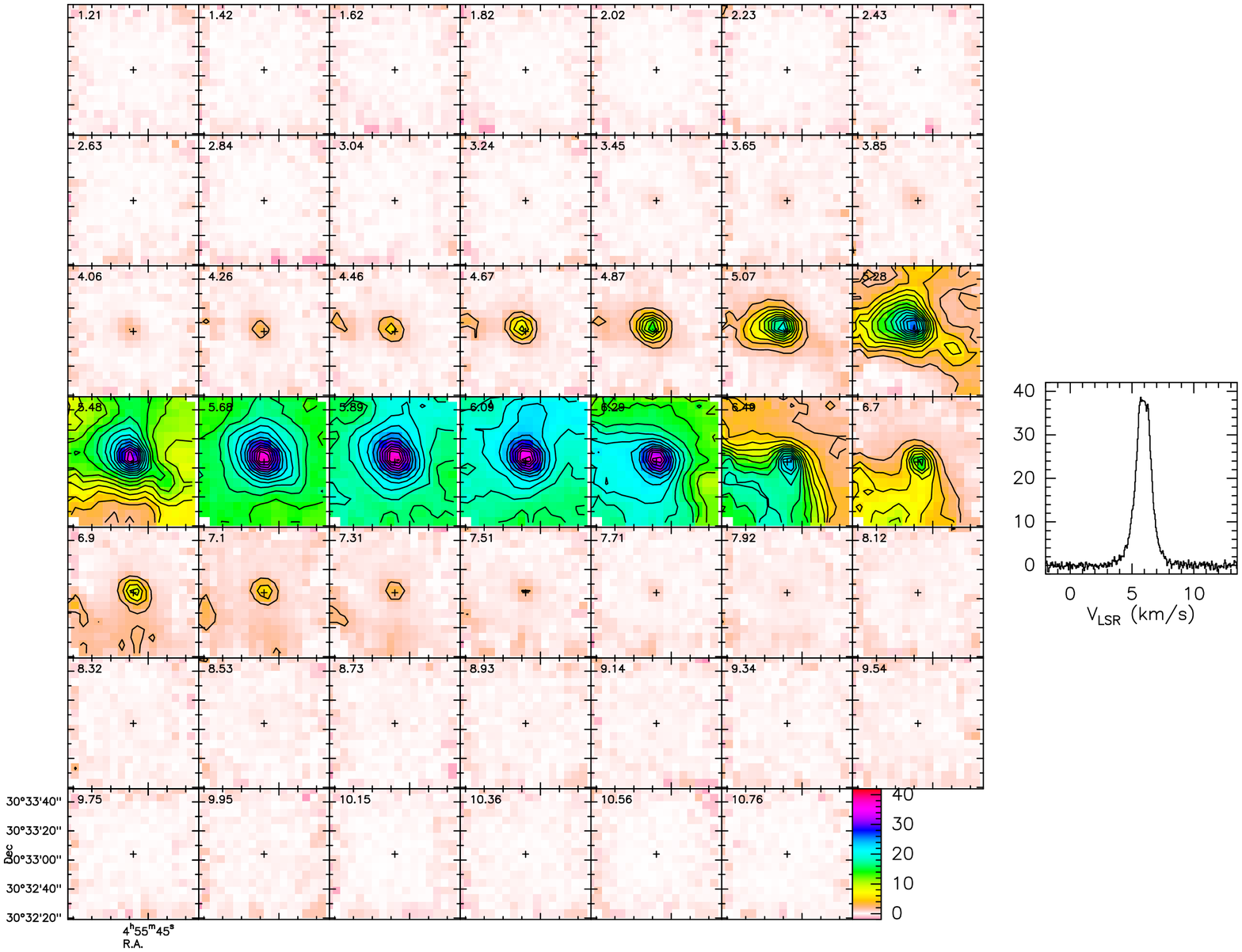} 
		\includegraphics[width=0.8\textwidth]{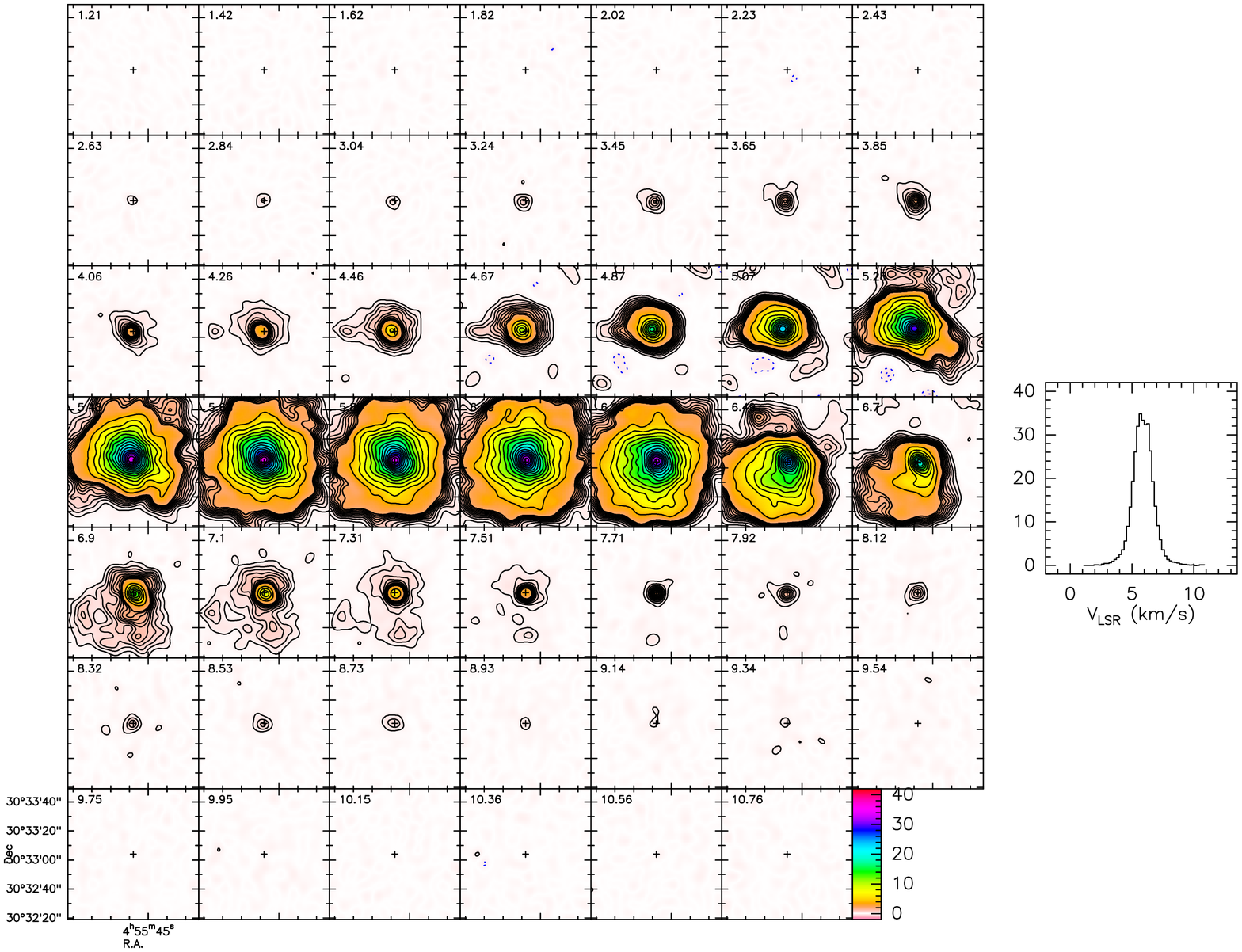} 
	\caption{Top panel: \dco~\jdu~channel maps from the 30-m observations at 11$\farcs$2 resolution. The field of view is 90$''$. The insert
shows the spectrum toward AB Aur at higher spectral resolution. Contour spacing is  2 K. Bottom panel:
combined \dco~30-m + SMA image at 7.4$''$ resolution. Contour levels are 1 to 5 K by 1 K, then
every 5 K (1 K = 0.42 Jy beam$^{-1}$ at this resolution). In the cleaning,
    the residual are around 0.2 Jy beam$^{-1}$ in the 16 inner channels.}
 	\label{fig:largeco}
\end{figure*}

\subsection{CO gas}
The channel maps and the spectrum obtained with the IRAM 30-m toward AB Aur are shown in the top panel in Fig. \ref{fig:largeco}. The angular resolution of the IRAM 30-m is 11$\farcs$2. The emission is extended and dominated by the envelope near the systemic velocity. There is a velocity gradient in the southeast-northwest direction. The combined image of the 30-m + SMA CO data with the smoothed resolution of 7$\farcs$4  is presented in the lower panel in Fig. \ref{fig:largeco}. The peak flux density is 84 Jy beam$^{-1}$, which corresponds to 40 K. At this resolution, the emission toward the south is clearly seen near the velocity of 6.9 to 7.3 km s$^{-1}$.

\subsubsection{Large-scale map at high angular resolution}
\label{sec:sub:co}

\begin{figure*}[!ht]
\begin{center}
\includegraphics[width=0.95\textwidth]{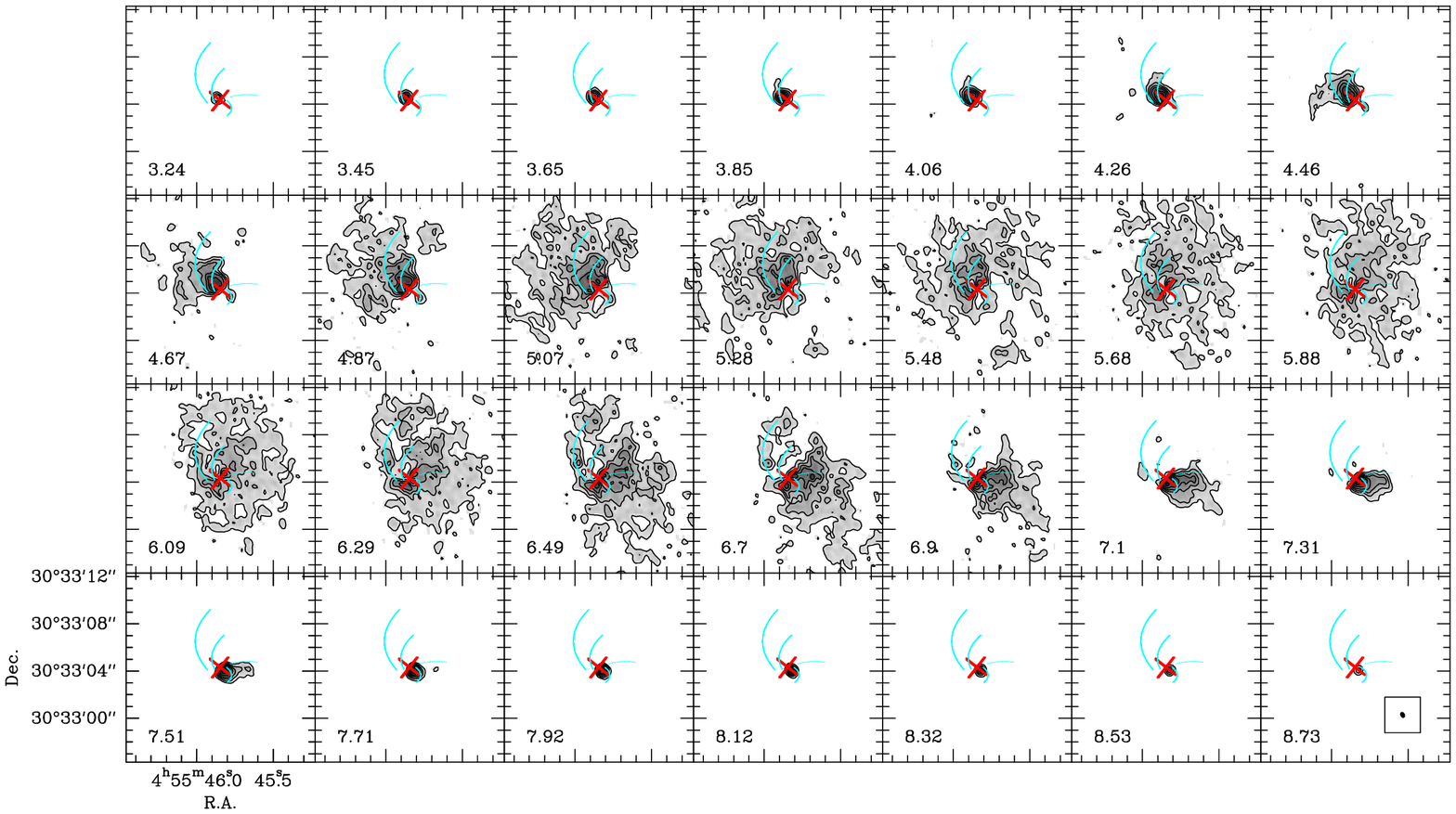}
\caption{Channel maps of the \dco~\jdu~line from the PdBI mosaic. The contours are 1 to 9 by 1 $\times$ 0.1 Jy per 0$\farcs$56 $\times$ 0$\farcs$42 beam (4.5$\sigma$, 9.91 K). The V$_{\rm LSR}$ is marked in each panel in km s$^{-1}$. The red cross marks the stellar location. Cyan arcs mark the CO spirals identified in Sect. 3.3.1. The S4 arm is a dotted arc, because it does not have as broad a linewidth as S1, S2, and S3.} \label{fig:cochannels}
\end{center}
\end{figure*}

We present in Fig. \ref{fig:cochannels} the \dco~\jdu~ mosaic image from PdBI.
The full data set was also studied, combining the 30-m data with the SMA field and the PdBI mosaic.
A comparison between the 30-m+SMA+PdB recombined image and the
PdB-only mosaic reveals that all the interesting detected structures are actually confined to within
the central beam of the PdBI mosaic, with few significant structures (at high resolution) beyond about 12$''$ from AB Aur. Moreover, the extended emission recovered
 in the 30-m+SMA+PdB image makes the structures less obvious. The PdB-only mosaic channel maps presented in Fig. \ref{fig:cochannels} are more suitable to pinpointing
the contrasts between different structures.
Since these images do not include the extended flux, the apparent brightness temperatures are lower limits, and correction for the
missing flux must, however, be included to derive physical parameters, such as temperature.

\begin{figure}[h!]
\includegraphics[angle=0,width=1\columnwidth]{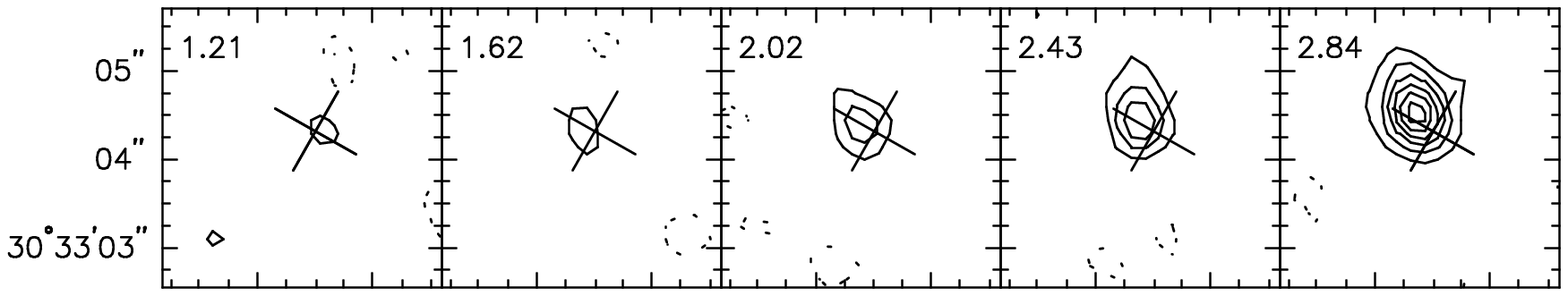}
\includegraphics[angle=0,width=1\columnwidth]{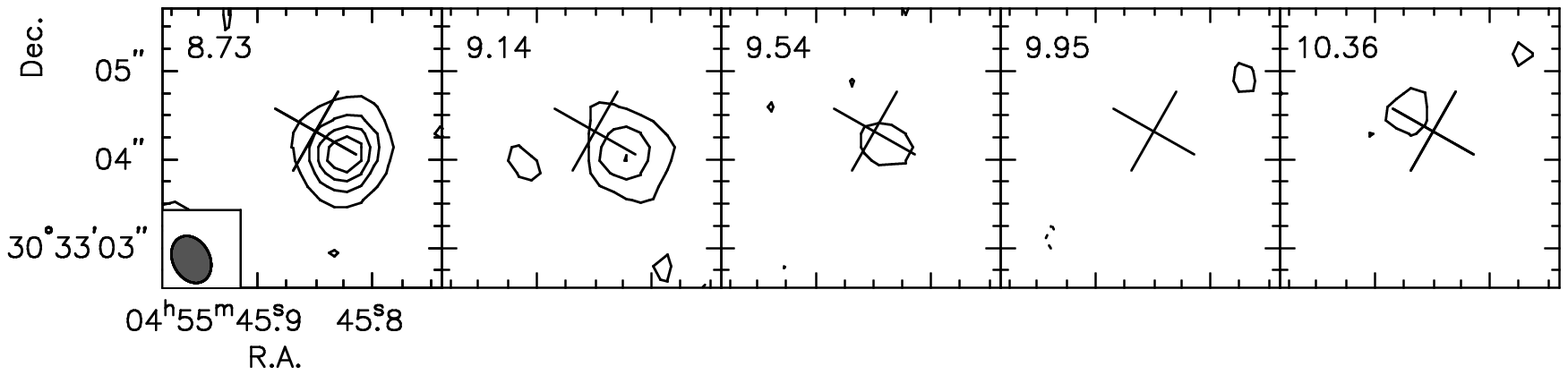}
\caption{The same as Fig. 3 but toward high-velocity channels.
The gray cross marks the size of the disk (145 AU) with a positional angle of 150$\degr$. The contours are multiples of 44 mJy per 0$\farcs$56 $\times$ 0$\farcs$42 beam.}
\label{fig:cohigh}
\end{figure}
\subsubsection{Inner gas}\label{sec:inner gas}

The \dco~\jdu~ emission is detected from the velocity of 1.6 km s$^{-1}$ to 9.5 km s$^{-1}$ (Figs. \ref{fig:cochannels} and \ref{fig:cohigh}).
The emission peaks in the high-velocity ends (Fig. \ref{fig:cohigh}) are near the 1.3 mm continuum central peak. The mean
position of the highest velocity CO peak is off by $\sim$ 0$\farcs05$ to the 1.3 mm stellar peak. Relative
positions are controlled by the bandpass calibration accuracy, which would yield relative positions valid
to about 1/100$^{\rm th}$ of the beamsize, and by the signal-to-noise. The difference is at 2 to 3 $\sigma$ level and is only marginally significant.

High-velocity CO gas (Fig. \ref{fig:cohigh}) is detected well within the dust ring, at least as close as 20 AU from the star.
This high-velocity emission is consistent with a rotating disk with the rotation axis of PA\,$= -38\degr \pm 1\degr$.
However, because of the restricted  velocity range (more than 2.4 km s$^{-1}$ away from the systemic velocity) and limited spatial extent,
the inclination angle is poorly constrained.

%
\begin{table}
\caption{Best-fit parameters of high velocity CO emission}
\begin{tabular}{l|l|l}
Fixed $i$ ($^\circ$) & [23] & [29] \\
\hline
Inner radius (AU) & $< 20$ &  $< 20$ \\
T (100 AU) (K) & $ 54 \pm 4$ &  $ 54 \pm 4$ \\
$q$ & $0.20 \pm 0.08$ & $0.14 \pm 0.07$ \\
PA ($^\circ$) & $-38.2 \pm 1.1$ &  $-38.2 \pm 1.1$ \\
V$_\mathrm{rot}$ (100 AU) (km.s$^{-1}$) & $5.60 \pm 0.15 $ & $4.57 \pm 0.12$ \\
$v$ & $0.50 \pm 0.04$ & $0.72 \pm 0.03$ \\
M$_*$ ($\msun$) & $3.5 \pm 0.15 $ & [$2.35 \pm 0.12$]\tablefootmark{*} \\
\end{tabular}
\tablefoot{Best-fit parameters are derived assuming two different inclination angles: i=23$\degr$ from best fit to dust emission, and i=29$\degr$ from stellar properties and Keplerian assumptions. $V_\mathrm{rot}$ is the
deprojected rotation velocity at 100 AU. Values in brackets are fixed.}
\tablefoottext{*}{Stellar mass only meaningful if $v = 0.5$.}
\label{tab:cohigh}
\end{table}
%
%

\begin{figure*}[!ht]
\begin{center}
\includegraphics[scale=0.8]{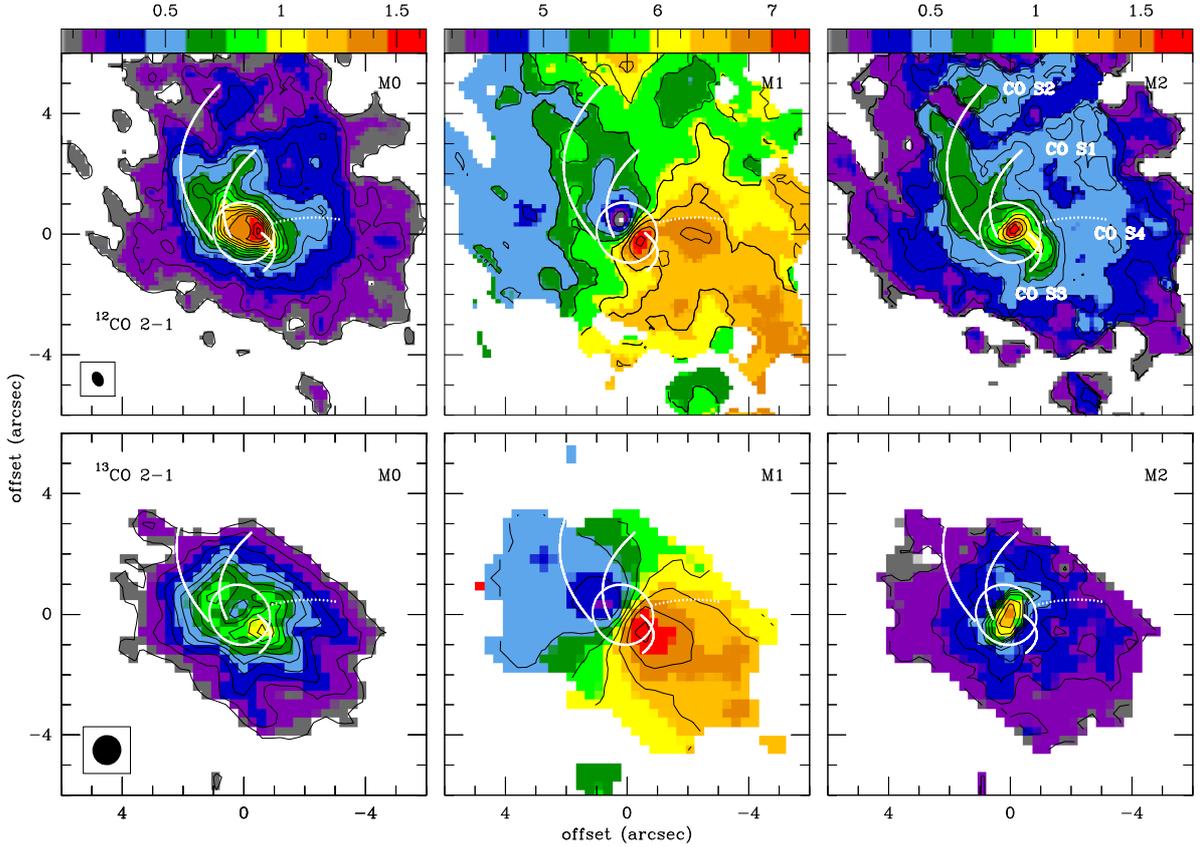}
\caption{Maps of moments 0, 1, and 2 from left to right panel for \dco~\jdu~(upper row; this work) and \tco~\jdu~\citep[lower row;][]{Pietu+etal_2005}. The white ellipse and arcs trace the best-fit dust ring and the identified CO spiral arms, respectively.
Left panels: The contours start from and step in 0.12 Jy beam$^{-1}$ km s$^{-1}$.
Middle panels: Contours steps are 0.4 km s$^{-1}$. Right panels: Contours steps are 0.1, 0.2, 0.3, 0.4, 0.6, 0.8, 1.0, ... km s$^{-1}$. The unit of the wedge is Jy beam$^{-1}$ km s$^{-1}$ for moment 0 maps and is km s$^{-1}$ for moment 1 and moment 2 maps.}
\label{fig:comoments}
\end{center}
\end{figure*}


Physical quantities based on two inclination angles are further derived and presented in Table \ref{tab:cohigh}.
We used a simple disk model and the DiskFit code \citep{Pietu+etal_2007} to derive them: the temperature, $T$, and velocity, $V$, are assumed to follow a power-law dependence on the radius, $r$: $T= T_0 (r/r_0)^{-q}$ and $V = V_0 (r/r_0)^{-v}$. On one hand,
if the inclination angle is fixed to $23^\circ$, corresponding to the best fit of the continuum ring, the best-fit velocity law will be
compatible with Keplerian rotation. The stellar mass (M$_{*}$) will be $3.5 \pm 0.2~\msun$, which is higher than the mass of 2.4 $\msun$ expected from the AB Aur spectral type. On the other hand, if the stellar mass is fixed to be 2.4 $\msun$ and the disk is assumed to be in Keplerian rotation, the best-fit inclination angle will be $29^\circ$,
which deviates from the best-fit value for the dust ring of $23^\circ$.
Finally, if the inclination angle is fixed to be 29$^\circ$, the best-fit velocity power-law index becomes $v = 0.72 \pm 0.03$, i.e., the Keplerian rotation assumption is rejected at the $7 \sigma$ level.
We do not find a common model that can fit both the dust ring and high-velocity CO emission.

In comparison, on a larger scale, the rotation axis as derived from $^{13}$CO by \citet{Pietu+etal_2005} has the PA = $-30\degr \pm 1^\circ$, which is consistent with the best-fit value of the dust ring. The inclination angle is in between $36^{\circ}$ and $42^\circ$, which is significantly larger than the values derived from our new high angular resolution data. We note that the velocity field on a large scale is sub-Keplerian. Another puzzling issue is the derived systemic velocity, $V_{\rm sys}$. It is $5.73 \pm 0.02$ km s$^{-1}$ as derived from the high-velocity parts only and is $5.85 \pm 0.02$ km s$^{-1}$ for the outer disk. This difference is marginally significant at the level of 4$\sigma$.
Such a deviation in $V_{\rm sys}$ may be due to an asymmetry in brightness between the
blue- and red-shifted wings, reflecting that the distribution of the gas is inhomogeneous.

\subsubsection{The disk}
The integrated intensity (moment 0), the intensity weighted velocity (moment 1), and the intensity weighted linewidth (moment 2) maps of \dco~\jdu~from this work and of \tco~\jdu~by \citet{Pietu+etal_2005} are shown in Fig. \ref{fig:comoments}. 
As a reference, a white solid ellipse that marks the 1.3 mm dust ring
is also indicated.
Figure \ref{fig:comoments} shows that \dco\ emission is
much more extended (7$''$ or 1000 AU) than \tco, which is strongly dominated by the rotating disk, and extends 
out to 4$''$ (560 AU). The outer radius in continuum is even smaller, at most 280 AU (see Fig. \ref{fig:dust_profile}).
While opacity differences naturally produce this ordering (R(dust) $<$ R(\tco) $<$ R(\dco)), the sharpness
of the dust profile would lead to only marginally larger radii for \tco\ and \dco\ if opacities only were to
explain this result. Our result thus confirms and reinforces the proposition made by \citet{Pietu+etal_2005} of
a change in dust properties (emissivity or gas-to-dust ratio) at a radius around 280 AU. Our more precise
determination is now in excellent agreement with the feature detected at 20 $\mu$m by \citet{Pantin+etal_2005}.
This suggests that a change in dust properties is more likely than a drop in gas-to-dust ratio beyond 280 AU.

\subsection{Spiral structure}\label{result:spiral}

%
\begin{figure}[!h]
\begin{center}
\includegraphics[width=0.8\columnwidth]{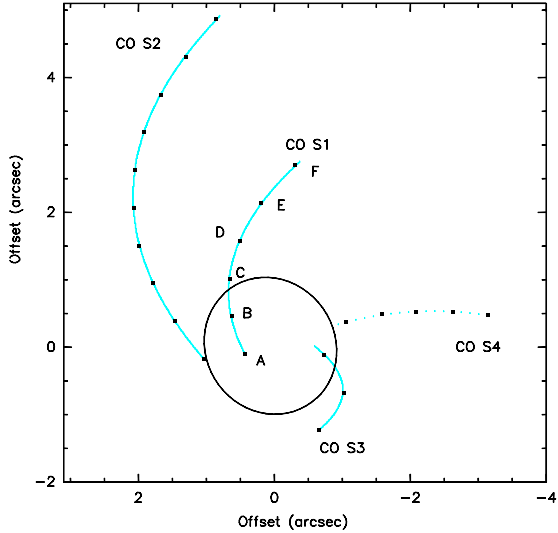}
 \caption{Locations on the spirals where the spectra in Fig. \ref{fig:spectra} are taken. The 1.3 mm dust ring is marked as a black ellipse.} \label{fig:id_spi}
\end{center}
\end{figure}
%
%

\begin{figure*}[!ht]
\begin{center}
\includegraphics[scale=1.0]{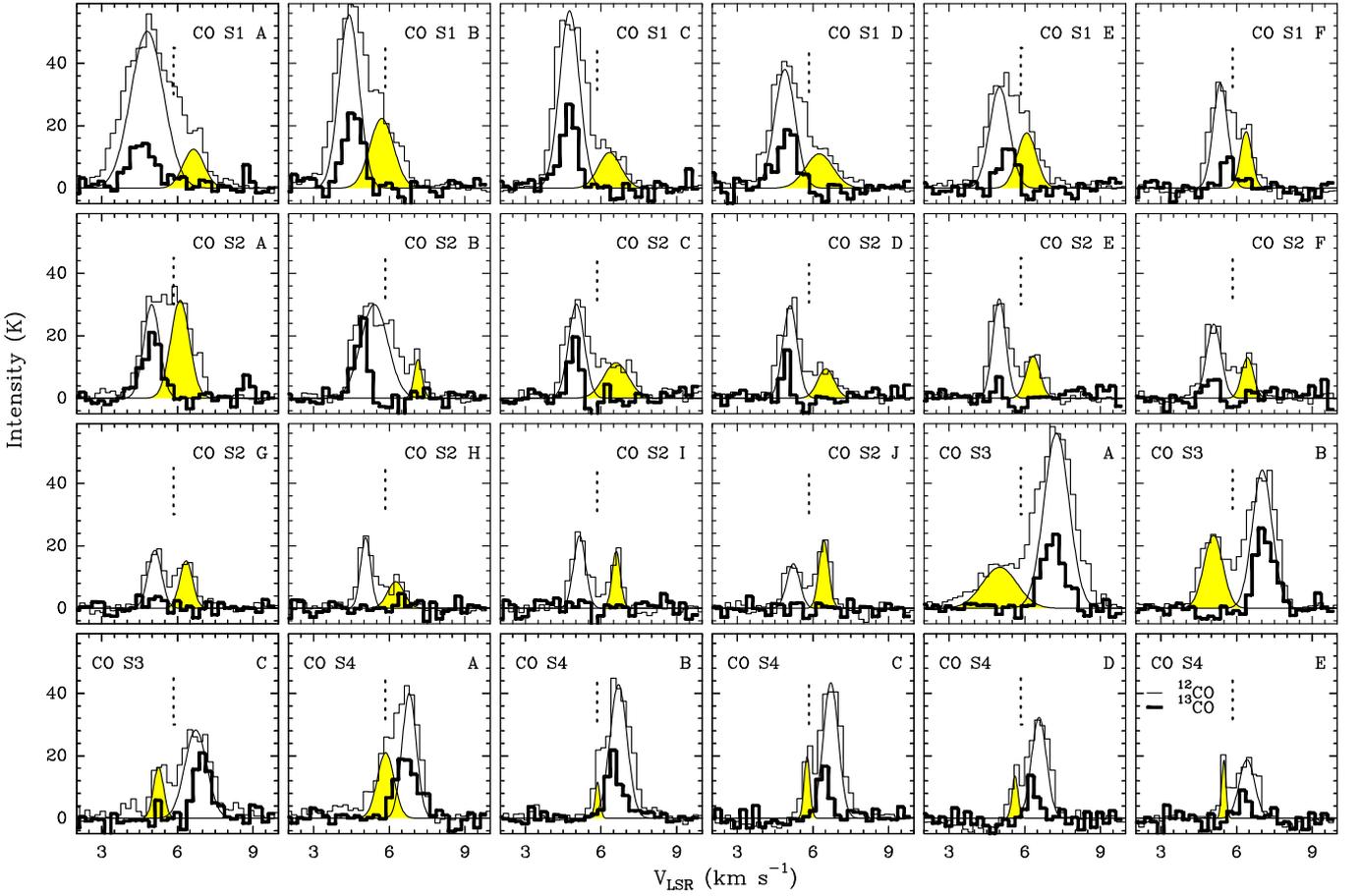}
\caption{Spectra of $^{12}$CO 2-1 (thin line) and $^{13}$CO 2-1 (thick line) on the locations of CO spirals. The alphabet order labels the distance to the stellar location: A is the closest, while J is the most distant one. The separations between the spectra is 0$\farcs$56. The shaded Gaussian indicates the integration area used to estimate the spiral mass. The dotted vertical line marks $V_{\rm sys}$ of 5.85 km s$^{-1}$.}
\label{fig:spectra}
\end{center}
\end{figure*}

\subsubsection{Identification of the spiral arms}
The \dco~moment maps exhibit very different morphology from those of \tco~\citep[from][]{Pietu+etal_2005}. While
    \tco~  traces the rotating disk, there are spiral-like patches with broader velocity dispersion and excess integrated intensity compared to the Keplerian patterns in the \dco~line.

The \dco~moment 0 and moment 2 maps from Fig. \ref{fig:comoments} allow us to identify four ``spiral''
patterns, CO S1 to CO S4.
From the moment 2 image, there are three patches where the linewidths are broader. We marked them as CO S1 to CO S3. Part of these arms can also be seen in the moment 0 map. Although the linewidth of CO S4 is not as broad as CO S1 to CO S3, there is excess emission at S4 in the moment 0 map. Besides the 0$\farcs$5 maps, we also checked the \dco~\jdu~maps with different tapering. The excess emission near the spirals can still be seen but is not as obvious as in the 0$\farcs$5 maps. On larger scales, the \dco~line might be too contaminated by the emission from the envelope in the combined maps. Another reason might be that the excess emission comes from structures that are actually very compact and not as bright as the Keplerian disk emission. We note that the labeled spirals are traced by first marking the locations where the moment 0 and moment 2 maps clearly show excess emission ``by eye''. Then, we fit four parabolic arcs to the marked locations.

We further checked the spiral emissions in the channel maps (Fig. \ref{fig:cochannels}).
The S1 and S2 arms are on the blue-shifted part of the disk. In S1, the spiral emission can be seen at 3.9 to 6.7 km s$^{-1}$.
The S2 arm is the most extended spiral among these four arms: it appears from 4.7 to 7.3 km s$^{-1}$ and connects to the SE part of the disk. 
The S3 and S4 arms are on the red-shifted part and at the near side of the disk.
The S3 arm has the broadest linewidth and it shows up from 4.3 to 7.3 km s$^{-1}$.
The S4 arm appears from 5.7 to 7.7 km s$^{-1}$. The bases of the four spiral arms are different. The S1 arm seems to trace back to the 1.3 mm gap (cavity; see the channels 3.9 to 5.5 km s$^{-1}$ in Fig. \ref{fig:cochannels} and moment 2 in Fig. \ref{fig:comoments}). The S3 arm can be traced
back to the ring inner edge (see channels 4.3 to 4.9 km s$^{-1}$ in Fig. \ref{fig:cochannels}). S2 and S4 arms originated somewhere near the dust ring.
The connection of these CO spirals with the dust ring and the NIR spirals and the comparison with the spirals detected in CO 3-2 will be discussed in Sect. \ref{sec:spiral_geometry}.

Spectra of \dco~\jdu~and \tco~\jdu~line along the spirals at locations spaced every 0$\farcs$56 (see Fig. \ref{fig:id_spi} for point labeling) are given in Fig. \ref{fig:spectra}. While the \tco~\jdu~line shows only the emission from the (nearly) Keplerian disk, the \dco~\jdu~spectra clearly exhibit a new kinematic component.
The \dco~\jdu~spectra can be decomposed into two Gaussians, 
one of which traces the rotating disk (at almost the same velocity as the \tco~\jdu~line) and the other traces an excess emission on the spiral.
They are named \dco$_{\rm disk}$~and \dco$_{\rm spiral}$, respectively.
Results of the Gaussian best-fit are listed in Table \ref{tab:fit_gauss}, and the kinematics are discussed in Sect. 4.2.3.
The velocity offsets of the two Gaussians are typically 1 to 2 km s$^{-1}$. This provides a natural explanation of the apparent broad linewidths in the spiral locations: there is a bulk of gas moving at different velocities from the rotating disk but not because of turbulent motions on the spirals. We note that although the spirals (arcs) are labeled by eye and not based on quantitative criteria, the parameters derived in this way are expected to reflect the bulk properties of these non-Keplerian component.

\begin{figure*}[!th]
	\centering
		\includegraphics[angle=0,width=0.9\textwidth]{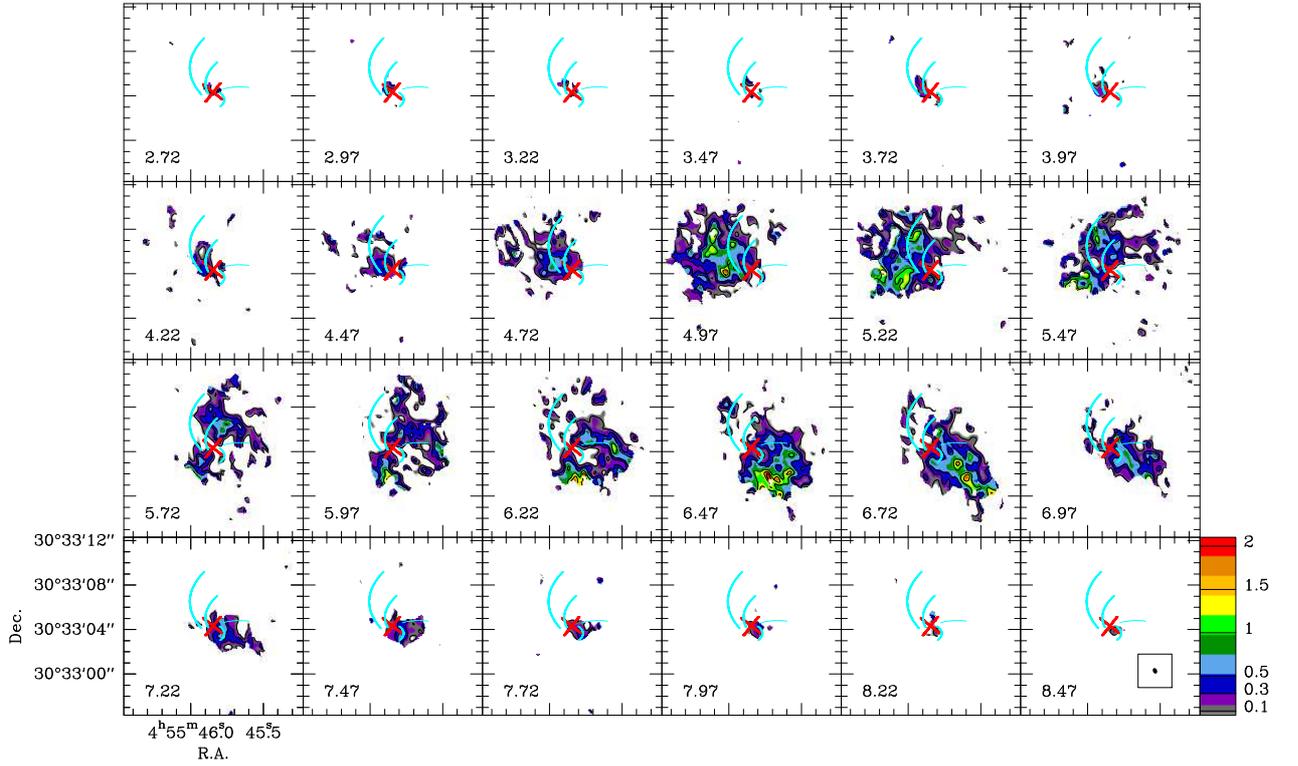} %
	\caption{The line ratio of \tco~\jdu~to \dco~\jdu~as a function of position in different velocities. There is some self-absorption (red color) in the southern parts. The \tco~emission otherwise has $< 1$ optical depth.}
 	\label{fig:iso-ratio}
\end{figure*}

\subsubsection{Temperature and opacity} 
The peak temperature in the CO maps, around 80 K, and the average brightness temperature in the inner 11$\arcsec$, 40~K, indicate
temperatures ranging from 30 K outside to 100 K in the inner regions of the disk, confirmed by the more sophisticated
disk analysis presented in Table \ref{tab:cohigh}.

The CO line opacity can be estimated from the $\tco/\dco$ ratio map (Fig. \ref{fig:iso-ratio}).
On large scales, in the velocity channels 5.22 to 6.72 km s$^{-1}$ in the southern part, and  4.97 to 5.47 km s$^{-1}$ in the northeast of the disk, the ratio $\tco/\dco$ is over 1 (Fig. \ref{fig:iso-ratio}). This must be due to self-absorption in \dco, implying optically thick in \tco~ with a temperature decreasing as a function of distance from AB Aur.

On the identified CO spirals, the $\tco/\dco$ ratio is low (in fact there is no robust \tco~detection), so the \dco~opacity
cannot be very high. A better constraint comes from the
apparent brightness temperature of the emission, which ranges from 10 K to 30 K. Assuming the kinetic temperature in the spiral is comparable to the one in the disk, this moderate brightness indicates opacities
of 0.3 to 1, since the spirals are resolved. 
Higher temperatures (hence lower opacities) cannot be  excluded, but this would require a multiline study.

\subsubsection{Gas mass and energetics inside the spirals}\label{section:gas_mass}
Under LTE conditions, the column density of the CO spirals can be estimated using the standard equation

\begin{equation}
N_{\rm CO} =  N_{\rm CO,J} \frac{Z}{\rm 2J+1} \rm exp\left(\frac{\rm h B J(J+1)}{\rm k T_{ex}}\right),
\end{equation}

\noindent where $N_{\rm CO,J}$ is the column density of the CO molecule in the quantum state J (J+1 $\rightarrow$ J). Here, N$_{\rm CO, J}$ is given by

\begin{equation}
N_{\rm CO, J} = 2.07 \times 10^3 \frac{2J+1}{2J+3} \frac{\nu^2}{A_{ul}} \int T_{\rm B} \rm{d}v,
\end{equation}
where B, $\nu$, $A_{ul}$, and $\rm T_{\rm B}$ dv are the rotational constant, the frequency in GHz, the spontaneous transition rate, and the brightness temperature T$_{\rm B}$ in K integrated over the linewidth dv in km s$^{-1}$, respectively.
Here, $Z$ is the partition function, $\sum_{\rm J=0}^{\infty}$ (2J+1) exp[-${\rm hBJ(J+1)/k\rm T_{\rm ex}}$], and T$_{\rm ex}$ is the excitation temperature.

We assume \rm T${_{\rm ex}}$= 50 K as
representative of the average temperature inside the arms. The derived column density of \dco~ and \tco~ in the spirals
are thus given by $N_{\rm \dco}$ = 8.9$\times$10$^{13}$ $\int T_{\rm B}$ dv and $N_{\rm \tco}$ = 6.07$\times$10$^{13}$ $\int T_{\rm B}$ dv, respectively.
With the integrated line flux calculated using $T_{\rm B}$ and $\triangle$v given in Table \ref{tab:fit_gauss}, the derived values of $N_{\rm ^{12}CO}$ range from 1.5$\times$10$^{14}$ to 1.5$\times$10$^{15}$ cm$^{-2}$. Using a CO/H$_2$ ratio of 10$^{-4}$, the total gas mass in these four spirals is only 2.5$\times$10$^{-7}$ M$_{\sun}$. Opacity effects could
increase these values.

An upper limit of the total mass of the spirals can be estimated from the \tco~\jdu~line, which is not detected in
the spirals. With the 3$\sigma$ noise level of 3 K per channel (0.25 km s$^{-1}$), $N_{\rm ^{13}CO}$ is less than 10$^{14}$ cm$^{-2}$, assuming the same linewidth as \dco~\jdu.
Assuming the $^{12}$C/$^{13}$C ratio 77 in the local ISM \citep{Wilson+etal_1994}, the upper limit of $N_{\rm \dco}$ and the gas mass in the spirals are $\sim$10$^{16}$ cm$^{-2}$ and 1.5$\times$10$^{-5}$ M$_{\sun}$, respectively.
In comparison, $N_{\rm \dco}$ in the disk at 100 AU estimated in \citet{Pietu+etal_2006} is 1.4$\times$10$^{19}$ cm$^{-2}$.
In the spirals, the column density is at least 10$^{3}$ times lower than the one in the disk.

%
%
\begin{table*}[th!]
\caption{Parameters of the Gaussian best-fit toward \tco~ and \dco}
\centering
\begin{tabular}{c|cc|c|ccc|ccc|ccc}
\hline
\multirow{3}{*}{ID} & \multicolumn{2}{c}{offset} & \multirow{3}{*}{$\theta_{\rm PA}$} &\multicolumn{3}{c}{\tco~} & \multicolumn{3}{c}{\dco$_{\rm disk}$} & \multicolumn{3}{c}{\dco$_{\rm spiral}$} \\ 
& R.A. & Dec. & & $V_{\rm LSR}$ & $T_{\rm B}$ & $\triangle$ v & $V_{\rm LSR}$ & $T_{\rm B}$ & $\triangle$ v & $V_{\rm LSR}$ & $T_{\rm B}$ & $\triangle$ v \\
 & ($\arcsec$) & ($\arcsec$) & ($\degr$) & (km s$^{-1}$) & (K) & (km s$^{-1}$) & (km s$^{-1}$) & (K) & (km s$^{-1}$) & (km s$^{-1}$) & (K) & (km s$^{-1}$)\\

  \hline
1A & 0.49 & -0.14 & -108 & 4.57$\pm$0.06 & 14.2 & 0.53 & 4.80$\pm$0.05 & 50.2 & 0.92 & 6.64$\pm$0.13 & 12.5 & 0.59 \\
1B & 0.73 & 0.42 & -180 & 4.55$\pm$0.03 & 24.8 & 0.36 & 4.40$\pm$0.05 & 55.5 & 0.56 & 5.70$\pm$0.18 & 22.3 & 0.67 \\
1C & 0.76 & 0.98 & 160 & 4.72$\pm$0.02 & 26.1 & 0.30 & 4.75$\pm$0.03 & 56.8 & 0.56 & 6.34$\pm$0.19 & 11.4 & 0.63 \\
1D & 0.58 & 1.54 & 147 & 4.98$\pm$0.04 & 18.0 & 0.37 & 4.89$\pm$0.06 & 38.0 & 0.58 & 6.25$\pm$0.36 & 11.0 & 0.77 \\
1E & 0.23 & 2.10 & 132 & 5.32$\pm$0.04 & 13.7 & 0.29 & [5.00] & 32.4 & 0.55 & 6.08$\pm$0.04 & 17.7 & 0.54 \\
1F & -0.36 & 2.66 & 114 & 5.63$\pm$0.04 & 10.6 & 0.21 & 5.36$\pm$0.03 & 33.6 & 0.40 & 6.38$\pm$0.05 & 18.1 & 0.33 \\\hline
2A & 1.20 & -0.19 & -135 & 4.97$\pm$0.04 & 19.4 & 0.38 & 4.98$\pm$0.03 & [30.0] & 0.44 & 6.11$\pm$0.04 & 31.3 & 0.53 \\
2B & 1.69 & 0.37 & -167 & 4.88$\pm$0.02 & 24.7 & 0.24 & 5.41$\pm$0.04 & [30.0] & 0.76 & 7.14$\pm$0.05 & 12.4 & 0.22 \\
2C & 2.07 & 0.93 & -177 & 4.94$\pm$0.03 & 20.1 & 0.20 & 5.02$\pm$0.03 & 30.1 & 0.44 & 6.59$\pm$0.11 & 11.3 & 0.69 \\
2D & 2.31 & 1.49 & 179 & 4.90$\pm$0.04 & 17.5 & 0.13 & 5.10$\pm$0.02 & 29.7 & 0.40 & 6.52$\pm$0.08 & 9.4 & 0.47 \\
2E & 2.40 & 2.05 & 177 & - & - & - & 4.99$\pm$0.01 & 31.9 & 0.36 & 6.33$\pm$0.04 & 13.5 & 0.39 \\
2F & 2.37 & 2.61 & 175 & - & - & - & 5.10$\pm$0.02 & 23.7 & 0.41 & 6.43$\pm$0.04 & 13.0 & 0.33 \\
2G & 2.22 & 3.17 & 172 & - & - & - & 5.10$\pm$0.02 & 18.5 & 0.38 & 6.34$\pm$0.03 & 15.2 & 0.36 \\
2H & 1.93 & 3.73 & 170 & - & - & - & 5.06$\pm$0.02 & 22.7 & 0.28 & 6.26$\pm$0.06 & 8.5 & 0.46 \\
2I & 1.52 & 4.29 & 166 & - & - & - & 5.15$\pm$0.02 & 23.1 & 0.34 & 6.60$\pm$0.02 & 18.2 & 0.23 \\
2J & 0.99 & 4.85 & 161 & - & - & - & 5.22$\pm$0.04 & 14.3 & 0.36 & 6.43$\pm$0.02 & 21.9 & 0.27 \\\hline
3A & -0.85 & -0.13 &  30 & 7.09$\pm$0.03 & 22.5 & 0.43 & 7.26$\pm$0.02 & 56.1 & 0.70 & 5.01$\pm$0.13 & 13.1 & 0.98 \\
3B & -1.18 & -0.69 &   1 & 7.08$\pm$0.03 & 25.2 & 0.36 & 7.02$\pm$0.02 & 44.2 & 0.55 & 5.08$\pm$0.03 & 23.4 & 0.53 \\
3C & -0.77 & -1.25 & -23 & 6.97$\pm$0.03 & 21.2 & 0.30 & 6.73$\pm$0.03 & 28.2 & 0.61 & 5.25$\pm$0.04 & 16.4 & 0.28 \\\hline
4A & -1.24 & 0.36 &  54 & 6.65$\pm$0.04 & 18.9 & 0.46 & 6.79$\pm$0.04 & [40.0] & 0.40 & 5.86$\pm$0.27 & 21.0 & 0.47 \\
4B & -1.85 & 0.47 &  39 & 6.47$\pm$0.02 & 20.0 & 0.28 & 6.69$\pm$0.02 & 42.9 & 0.48 & 5.86$\pm$0.03 & 11.6 & 0.14 \\
4C & -2.44 & 0.51 &  29 & 6.40$\pm$0.03 & 17.3 & 0.22 & 6.71$\pm$0.02 & 43.4 & 0.42 & 5.77$\pm$0.02 & 19.9 & 0.16 \\
4D & -3.05 & 0.51 &  22 & 6.27$\pm$0.04 & 13.9 & 0.17 & 6.57$\pm$0.02 & 32.3 & 0.42 & 5.62$\pm$0.03 & 13.6 & 0.16 \\
4E & -3.64 & 0.46 &  17 & 6.28$\pm$0.05 & 9.6 & 0.17 & 6.41$\pm$0.03 & 18.8 & 0.43 & 5.49$\pm$0.06 & 18.6 & 0.12 \\

\hline
\end{tabular}
\label{tab:fit_gauss}
\tablefoot{Best-fit results of the spectra shown in Fig. \ref{fig:spectra}. Coordinates of the spectra are listed in offset with respect to (R.A., Dec.)$_{\rm J2000}$ = (04$^h$55$^m$45$\farcs$84, +30$\degr$33$\arcmin$04$\farcs$30). $V_{\rm LSR}$, $T_{\rm B}$ and $\triangle \rm v$ is the observed velocity with respect to the local standard of rest, the brightness temperature and the line-width, respectively. $\theta_{\rm PA}$ is the position angle of the offset location with respect to the major axis of the disk (-121.3$\degr$ from north), assuming an inclination of 23$\degr$. Values in brackets are fixed. Note that there is no robust \tco~ detection at locations CO S2E to CO S2J. \\
}
\end{table*}
%
%
%


The kinetic energy contained in the spirals in the rotating disk frame are calculated using the equation
\begin{equation}
 E_{\rm kin}=0.5{M}_{\rm gas}v_{\rm off}^{2},
\end{equation}
where $v_{\rm off}$ is the difference between the $V_{\rm LSR}$ of the \dcos~component and the $V_{\rm LSR}$ of the \tco~component listed in Table \ref{tab:fit_gauss}.
The resulting $E_{\rm kin}$ on the spiral is 6$\times$10$^{36}$ erg. The upper limit of $E_{\rm kin}$ given from \tco~\jdu~ is 3$\times$10$^{38}$ erg, assuming the same $v_{\rm off}$ as the \dcos~component with the estimated upper limit of the spiral mass.

If the gas on the spirals mainly moves on the disk plane, the reported energy will be a factor of six higher, assuming that the inclination angle is 23$\degr$.
In comparison, the binding energy of a companion with a mass of 0.03 $\rm M_{\sun}$ at 45 AU, is 2.4$\times$10$^{43}$ erg, which is at least four orders of magnitude greater than the kinetic energy contained in the spirals. A small perturbation of the orbit of such a companion is sufficient to release enough energy to sustain such spirals. If emission is at a larger $h$ and is optically thick, the estimated $\rm E_{kin}$ will be the lower limit. The angular momentum is estimated to be 10$^{47}$ in cgs following the same method.

\section{Discussion}

\begin{figure*}[th!]
\includegraphics[scale=0.9]{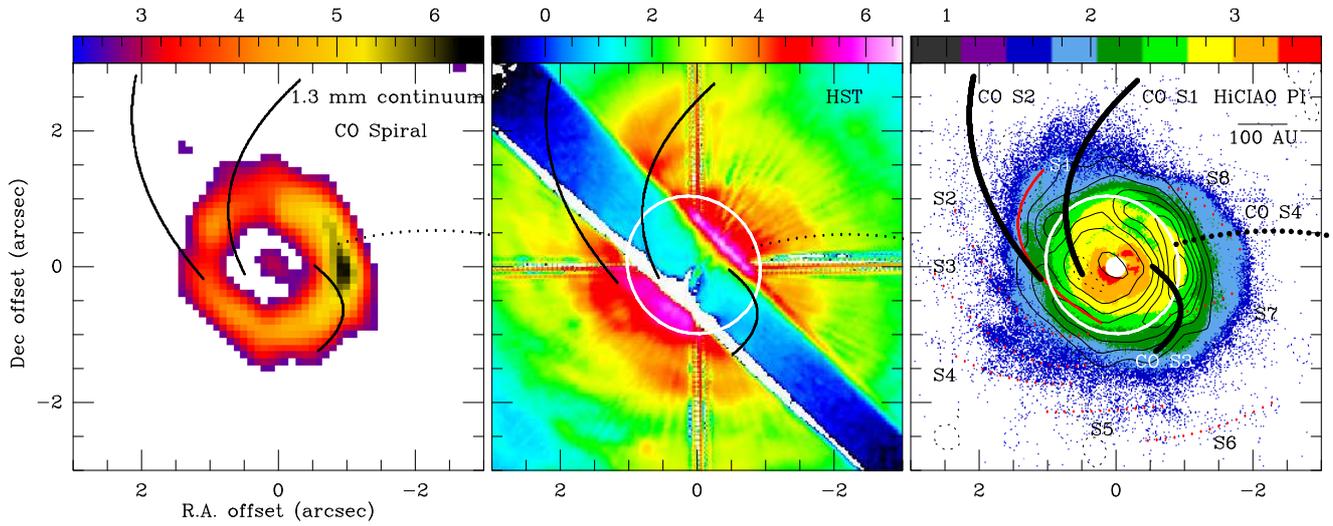}
{\caption{Left panel: dust continuum at 1.3 mm (color scale) and CO spirals (black arcs). Middle panel: HST image (color scale) and CO spirals. The 1.3 mm dust ring is marked as a white ellipse. Right panel: Comparison of 1.3 mm continuum (contours) and the HiCIAO polarization intensity (PI) image (color scale) at NIR. The black and red arcs mark
the CO spirals identified in Fig. \ref{fig:comoments} and the HiCIAO arm \citep{Hashimoto+etal_2011}, respectively.
The S1 arm is marked as a solid arc, because it is the one with apparent corresponding CO spirals.
The coronagraphic occulting mask of 0$\farcs$3 diameter is marked as a solid-white circle.}
\label{fig:over_spi}}
\end{figure*}

At high angular resolution both in the NIR and the submm domains, the surroundings of AB Aur are complex.
Figure \ref{fig:over_spi} is a
montage of the 1.3 mm continuum image with CO spirals, the optical image from HST, and the NIR image from HiCIAO of Subaru telescope. Superposed are the spirals as determined from the NIR by \citet{Hashimoto+etal_2011} and the CO spirals (this work).
The HST image from \citet{Grady+etal_1999} on large scales also reveal the spiral arms, which were later reported closer to the central star by \citet{Fukagawa+etal_2004} from Subaru deep images.

In dust emission, the large inner cavity and well-defined dust ring appear much like a scaled-down version of the tidal
cavity+ring observed in the binary system GG Tau/A \citep{Guilloteau+etal_1999}. More recently, several SMA, CARMA, and IRAM PdBI
observations have revealed the existence of such large cavities inside disks, such as HD 135344B, MWC\,758, LkCa 15, or even HH\,30. The material around AB Aur also differs from similar objects in some important aspects:
\begin{itemize}
\item In the large-scale CO emission, four spiral-like arms are detected. Part of these CO spirals are also seen in NIR even though the existence of a
remaining envelope of dust, which affects NIR image mostly, and gas (\dco) partially contaminates
any detailed analysis of the material close to the star.
\item The disk is not in Keplerian rotation, and the dust ring is clearly asymmetric both in continuum and in the \dco~and~\tco~lines.
\item The inner cavity is not fully devoid of gas. This gas is likely not uniform within the central hole (see Sect.\ref{sec:inner gas}).

\item Fitting the large-scale disk traced in \tco~and \cdo~(see Sect. \ref{sec:inclination}) and the dust ring separately leads to some apparent inconsistencies. In particular the inclination angles differ from 42$\degr$ to 23$\degr$ (while the typical errors are
    1$^\circ$ to 3$^\circ$).
\end{itemize}

In this section, we discuss these points and see how they can be interpreted and possibly
reconciled in a single picture.
We focus first on a comparison of the mm/submm data with other similarly resolved observations and
on the ring properties. We then discuss an overall scheme for the spiral features and present the
dynamics of the AB Aur system.

\subsection{Properties of the ring and inner material}

\subsubsection{Dust gap and inner material}
The 1.3 mm continuum peak at the stellar location might suggest the existence of, at least, one inner disk.
The lower limit of the total gas+dust mass at the stellar location is estimated to be 4$\times$10$^{-5}~\msun$ (Sect. \ref{sec:high angular resol}).
Although we cannot rule out the possibility that the 1.3 mm stellar peak comes from the optically thick free-free emission, an inner disk has been detected and resolved in the near infrared (NIR) and mid-infrared (MID-IR) interferometric observations by \citet{Eisner+etal_2004} and \citet{diFolco+etal_2009}.
The inner disk is also required to explain the observed accretion rate of 10$^{-9}$ to 3$\times$10$^{-7}\msun$ yr$^{-1}$ \citep{Brittain+etal_2007,Telleschi+etal_2007,Donehew+etal_2011}.

Besides the continuum emission at the stellar location, the detection of ``high'' velocity \dco~\jdu~emission inside the dust cavity also indicates that this area is not devoid of gas.
The gas mass estimated from the CO 2-1 line with velocity more than 2.5 km s$^{-1}$ away from $V_{\rm sys}$ is $\sim$10$^{-5}$ $\msun$. These gases are 50 AU away from the stellar location with an orbital period of 200 yr. From the orbital timescale argument, which the mass accretion rate toward the center cannot exceed the orbiting mass divided by the orbiting period, the upper limit of the accretion rate is in the order of 10$^{-7}$ ${\rm \msun~yr^{-1}}$, consistent with the values reported in previous paragraph.

\subsubsection{Ring properties}\label{discuss:ring}

Comparing the 1.3 mm continuum with NIR emission is not straightforward, because the dust disk has very high optical depth at NIR.
The dust emission at these two wavelengths originates in different dust layers.
The NIR emission traces the scattered-stellar light by dust grains in the upper layers of the flared disk, while the 1.3\,mm thermal emission, which is essentially optically thin, characterizes the bulk of the disk mass.
Although the very inner parts of the NIR maps are occulted (e.g HiCIAO data: the
area of the mask is seen in white in Fig. \ref{fig:over_spi}), the superposition of the mm and infrared images reveals coherent behavior (Fig. 10).
We find that the NIR emission peaks before the mm ring at radius around 70 AU, then the mm emission peaks at radius 150 AU followed by a
decrease in the NIR brightness, indicating that there are fewer stellar photons impinging on the disk behind
the bulk of the dust emission. This behavior is fully consistent with predictions from stellar irradiated-dust disk models \citep[see for example,][]{Dullemond+etal_2007a}.

\begin{figure*}[ht!]
\includegraphics[scale=0.9]{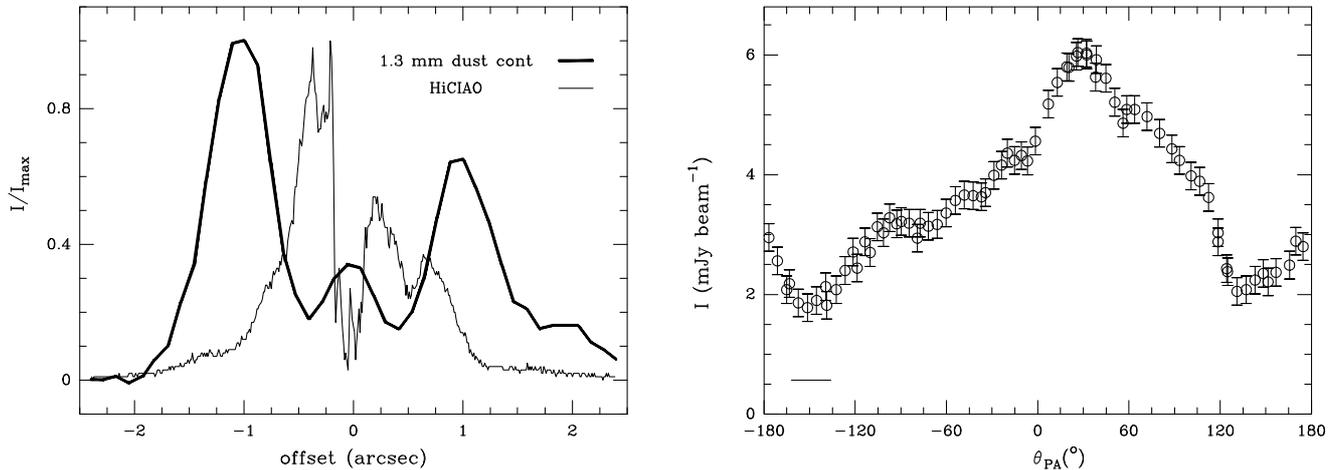}
 \caption{Left panel: plot of the normalized intensity of 1.3 mm continuum and HiCIAO PI image at NIR along the major axis of the disk's major axis. The central 0$\farcs$3 in HiCIAO data is occulted. Right panel: Plot of 1.3 mm intensity versus $\theta_{\rm PA}$ detected with the PdBI data of the best fit dust ring, marked as a white ellipse in Fig. \ref{fig:continuum}, bottom right panel.
The error bar is set to 0.23 mJy beam$^{-1}$. The center of the ellipse is on the 1.3 mm stellar peak. The horizontal bar at the lower left corner marks the synthesized beam (0$\farcs$45).
 } \label{fig:azimuth}
\end{figure*}

The AB Aur disk is not azimuthally homogeneous.
The best-fit models derived from the 1.3 mm and the \dco~ data in Tables \ref{tab:cont} and \ref{tab:cohigh} assume uniform density distribution with azimuth. The residual image of the 1.3 mm continuum (i.e. after subtracting the best-fit ring; lower middle panel of Fig. \ref{fig:continuum}) clearly shows the inhomogeneity in both radius and azimuth.
The azimuthal asymmetry can be quantified by Fig. \ref{fig:azimuth} where we trace the dust intensity at 1.3\,mm versus the position angle with respect to the major axis of the disk, $\theta_{\rm PA}$ (measured in the disk plane, by assuming an inclination of 23 $\degr$), inside the ring (peak location, Fig. \ref{fig:continuum}).
The major axis of the disk is assumed to be -121.3$\degr$ from north, since the disk rotation axis is at -31.3$\degr$ from the best fit of \tco~ in \citet{Pietu+etal_2005}.
There are two dips in the ring, one at $\theta_{\rm PA}$ of 130$\degr$ and the other at -150$\degr$ (Fig. \ref{fig:azimuth}).
The emission peak is at $\theta_{\rm PA}$ of about 30$\degr$.
The maximum intensity contrast on the ring is about 3.

\subsection{CO disc}

\subsubsection{Inclination angle}\label{sec:inclination}
The determined inclination angle $i$ of AB Aur disk varies with scale.
For the inner disk (radius $\ge$ 20 AU), $i$ is 23$\degr$ to 29$\degr$ based on the high velocity \dco~\jdu~line (Sect. 3.2.2; Table 2).
With the best-fit dust ring at 1.3 mm (radius of 145 AU), $i$ is 22$\degr$ to 24$\degr$ (Sect. 3.1.2; Table 1), consistent with the inclination derived from the high-velocity \dco~line.
We note that the derived inclination angle from the \dco~\jdu~data is based on the high-velocity components (2.4 km s$^{-1}$ offset from the $V_{\rm sys}$), which are less likely to be contaminated by the envelope.
On a larger scale (radius 70 AU to 600 AU), \citet{Pietu+etal_2005} reports $i$ to be $36\degr$ based on the \cdo~\juz~line.
Because the \cdo~\juz~line is optically thin and also less likely to be contaminated by the large-scale envelope, the discrepancy between the values derived from \dco~and \cdo~suggests a physical warp.
We note that the low $i$ toward a smaller scale has already been reported.
Based on the NIR interferometric data, \citet{Eisner+etal_2004} find that $i$ is $\le 20\degr$ for an inner disk with a radius less than 1 AU.
The $i$ inferred from our new 1.3 mm continuum and CO data is more consistent with  the value derived from the NIR.

Warped disks have been reported in the literature, e.g., Beta Pictoris \citep[and references therein]{Golimowski+etal_2006}.
The AB Aur system is somewhat similar to the GG Tau A system.
\citet{Guilloteau+Dutrey_2001} show the high angular resolution
\dco~\jdu~observations of CO gas inside the cavity with slightly different inclination angle  from the one derived from the dust ring.

\subsubsection{Rotation}
The rotation velocity also varies with scale.
In \citet{Pietu+etal_2005} and \citet{Lin+etal_2006}, the CO emission
was found to be in non-Keplerian rotation. This is partly due to the confusion with the remaining envelope since the \dco~\jdu line and especially the \dco~\jtd line are mostly optically thick, tracing both the envelope and the disk.
However, observed with \tco~and \cdo, where confusion is negligible, \citet{Pietu+etal_2005} measured the velocity power-law exponents $v$ of 0.42, 0.37, 0.47, and 0.82 for \tco~\jdu, \tco~\juz, \cdo~\jdu, and \dco~\jdu, respectively.
These exponents were fitted with all the detected emission in \citet{Pietu+etal_2005}, which is probably dominated by the larger scale disk.
The new \dco~\jdu~data at higher angular resolution add an additional puzzle, since the velocities within 100 AU appear to exceed the orbital velocity around a $\sim$2.4 $\msun$ mass star. One might explain such high velocities with an increasing inclination angle at smaller radius, but the implied disk warp is opposite to the one discussed in
Sect.\ref{sec:inclination} above. Given the possible brightness asymmetry  (see Sect.\ref{sec:inner gas}), a better explanation
is perhaps that the interpretation in terms of Keplerian disk is totally inappropriate. Instead, this gas may be tracing streamers of gas infalling (and rotating) from the ring onto the central object.

\subsection{CO spirals}

Our observations show that the disk structure in AB Aur is highly asymmetric.
Although sensitive CO 3-2 and CO isotopes are still needed to better constrain the temperature and density on the spirals, our data provide the dynamics of this morphological component for the first time.

\subsubsection{Geometry}\label{sec:spiral_geometry}

Spiral-like structures have been detected both in the CO gas and the dust emission/scattering light at various wavelengths, although they do not come from exactly the same locations.
In the IR, \citet{Fukagawa+etal_2004} report four spirals (one inner arm,
two outer arms with a branch on one of the outer arms).
More recently, \citet{Hashimoto+etal_2011} identified eight spirals, namely S1, S2, ..., S8, based on the infrared polarization-intensity image.
In comparison, the CO S2 shares part of the base with the infrared S1 (inner arm in Fukagawa's spiral) and S2 arms, which are in the southern part of the 1.3 mm dust ring (Fig. \ref{fig:over_spi}).
The CO S3 arm seems to trace the inner part of infrared S3 sprial (Fukagawa's outer arm).
We note that the spiral detected at 5.68 km s$^{-1}$ in CO 3-2 by \citet{Lin+etal_2006} appears in between CO S1 and CO S2 and traces the NIR S1 spiral more, suggesting that it is at a more upper layer than the CO 2-1 spirals.

In NIR, the spirals located in the southern region appear shorter, suggesting an inclination effect on a geometrically flaring structure. The spirals closer to us are compressed compared to those located on
the other side. This is also consistent with the NIR data, which reveal that the southern part is brighter and thus closer (forward scattering).
To constrain the physical conditions of the spirals, high angular-resolution and high-sensitivity observations of $\rm \dco~ and~\tco$ at multiple J states are needed.

\subsubsection{Gas kinematics}

So far, we only have the velocity information along the line of sight. Assumptions about the 3-D geometry are required to recover further information. Starting from the disk geometry, the observed velocity of gas motions with respect to the local standard of rest (LSR), $V_{\rm LSR}$, can be decomposed in cylinder coordinate ($r$, $\theta_{\rm PA}$,$z$) :
\begin{eqnarray}
V_{\rm LSR} & = & V_{\rm sys} + [V_{\rm rot}(r){\rm cos}(\theta_{\rm PA}) + V_{\rm rad}(r){\rm sin}(\theta_{\rm PA})]{\rm sin}(i) \nonumber \\
 & +&  V_{\rm z}(r) {\rm cos}(i), \label{eq:vlsr}
\end{eqnarray}
where $V_{\rm sys}$ is the systemic velocity with respect to LSR. We chose 5.85 km s$^{-1}$ for AB Aur in this calculation, because the spiral-like structures are extended.
Here, $V_{\rm rot}$ is the rotation velocity (positive in counter-clockwise), while $V_{\rm rad}$ (positive for outward motions) is the radial velocity in the disk plane, and $V_{z}$ is the component perpendicular to the disk plane.
The symbol $\theta_{\rm PA}$ is the position angle of each location with respect to the major axis of the disk (-121.3$\degr$ from the north),
and $i$ is the inclination angle of the disk axis with respect to the line of sight.
When calculating $\theta_{\rm PA}$ in each location of spirals, we assume $i$=23$\degr$, because this best-fit model gives a Keplerian disk (see Table \ref{tab:cohigh}). 

%
%
%
\begin{figure}[h!]
\includegraphics[width=0.7\columnwidth,angle=-90]{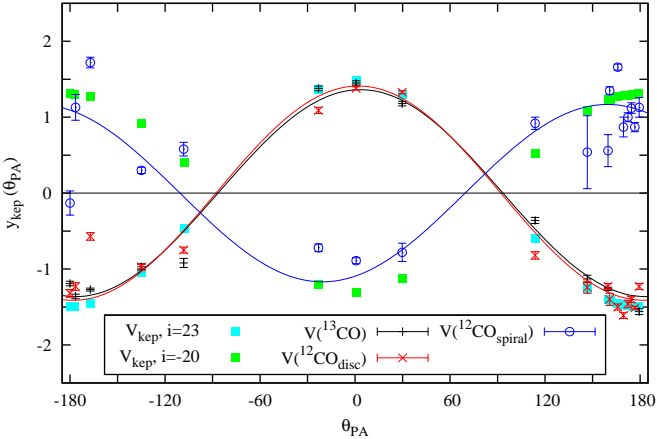}
 \caption{Plot of $\theta_{\rm PA}$ versus y$_{\rm kep}$ ($\theta_{\rm PA}$) as defined in Eq. \ref{eq:ypa_kep} of $\rm ^{13}CO, ^{12}CO_{\rm disk}~, ~ ^{12}CO_{spiral}$ and a purely Keplerian rotation in corresponding colors. The best-fit of each component is shown as a curve.} \label{fig:spa_v_3spec}
\end{figure}
In this decomposition, motions perpendicular to the disk plane would not depend on $\theta_{\rm PA}$. The symmetric motions perpendicular to the disk midplane cannot be seen: these would only add up velocity dispersion after considering the limited angular resolution, 
but not change the projected mean velocity.

For pure Keplerian rotation, $V_{\rm rot}$ equals $\sqrt{GM/r}$, where $M$ is the enclosed mass within a radius $r$, and $V_{\rm rad}$ and V$_{z}$ are 0.  Equation \ref{eq:vlsr} can be rearranged as
\begin{equation}
y_{\rm kep}(\theta_{\rm PA})\equiv(V_{\rm LSR}-V_{\rm sys})\sqrt{r} = \sqrt{GM}{\rm cos} (\theta_{\rm PA}) {\rm sin}(i), \label{eq:ypa_kep}
\end{equation}
where y$_{\rm kep}$($\theta_{\rm PA}$) now only depends on cos($\theta_{\rm PA}$).\\

Similarly, if we can correct for radial dependencies, pure radial motions would vary as sin($\theta_{\rm PA}$). For example, for free-fall accretion in the disk plane, $V_{\rm rad}$ equals $\sqrt{2GM/r}$,  $V_{\rm rot}$ and V$_{z}$ are 0.  Equation \ref{eq:vlsr} can be rearranged as
\begin{equation}
y_{\rm inf}(\theta_{\rm PA})\equiv(V_{\rm LSR}-V_{\rm sys})\sqrt{r} = - \sqrt{2GM}{\rm sin} (\theta_{\rm PA}) {\rm sin}(i). \label{eq:ypa_inf}
\end{equation}
Thus, a display of $y = [V_{\rm LSR}-V_{\rm sys}] \sqrt{r}$ as a function of $\theta_{\rm PA}$ can reveal the relative contributions of (Keplerian)
rotation and other motions. We fitted 1) a pure rotation $y = a_1 \cos(\theta_{\rm PA} - a_2)$, and 2) a combination
 of rotation and radial motions $y = a_3 \cos(\theta_{\rm PA}) + a_4 \sin(\theta_{\rm PA})$ to the observed velocities in the spirals.
We excluded Location 1E from the fit, because the two velocity components are not separated well here.
We also exclude CO S4 in the fitting, because there is less excess emission on the spirals. The fitting results of a pure rotation are shown in Fig. \ref{fig:spa_v_3spec} and in Table \ref{table:fit_ypa}.
For illustration, a pure Keplerian rotation with $M$=2.4 M$_{\sun}$ (based on the spectral type of AB Aur) and $i$=23$\degr$ (based on the axis ratio of the dust ring) is also shown.
We note that this choice of a representable Keplerian rotation in AB Aur is not trivial, because we do not find a consistent model based on our 1.3 mm continuum map and the high-velocity \dco~\jdu~line (see Sect. \ref{sec:high angular resol} and \ref{sec:inner gas}).
Nevertheless, the chosen $M$ of 2.4 M$_{\sun}$ and $i$ of 23$\degr$ can describe the general trend of the derived $y$ values of the \dco~{\rm disk} component and also the \tco~line.  
We also plot a Keplerian rotation with $M$=2.4 M$_{\sun}$ but with $i$=-20$\degr$ in Fig. \ref{fig:spa_v_3spec} in order to demonstrate that the \dco~$_{\rm spiral}$ component can originate in corotating gas with high inclination angle with respect to the disk plane. See section \ref{sec:global_picture} for more discussion.

Despite the previously noted apparent deviations from Keplerian motions \citep[best-fit velocity exponent  0.42,][]{Pietu+etal_2005}, the disk component is very close to pure Keplerian rotation, because the best-fit value $a_2$ of $V_{\rm \tco}$ and $V_{\rm \dco_{disk}}$ are 3$\degr\pm$4$\degr$ and 2$\degr\pm$5$\degr$, respectively.
The kinematics of the spiral component shows more deviations from a simple cosine function.
To the first order, it is fitted with
a similar $a_2 \simeq 0$ but with negative velocity. In other words, the kinematics of the spiral is apparently dominated by rotation
\textit{in the opposite direction to the disk spin}, with little contribution from radial motions.  This is confirmed
by the fit of rotation + radial motions. The best fit gives $a_3$ of $-1.09\pm0.10$, which is $\simeq a_1$, and $a_4$ of 0.41$\pm$0.22, which is only marginally different from 0. That $a_4$ is 2.5 times smaller than $a_3$ suggests that the motions within the spirals are mainly rotation with slightly radial motions.

%
\begin{table}[!ht]
 \caption{Best fit of y$_{\rm kep}$($\theta_{\rm PA}$)}              
 \label{table:fit_ypa}      
 \centering
\begin{tabular}{c |c c c}

   & $\tco$ & $\dcod$ & $\dcos$ \\
  \hline
 $a_1$ & 1.36 $\pm$ 0.04 & 1.41 $\pm$ 0.05 & -1.17 $\pm$ 0.12 \\
 $a_2$ & 3 $\pm$ 4 & 2 $\pm$ 5 & -21 $\pm$ 10 \\
\end{tabular}
 \end{table}

%
%
\begin{figure}[h]
\begin{center}
\includegraphics[width=0.8\columnwidth]{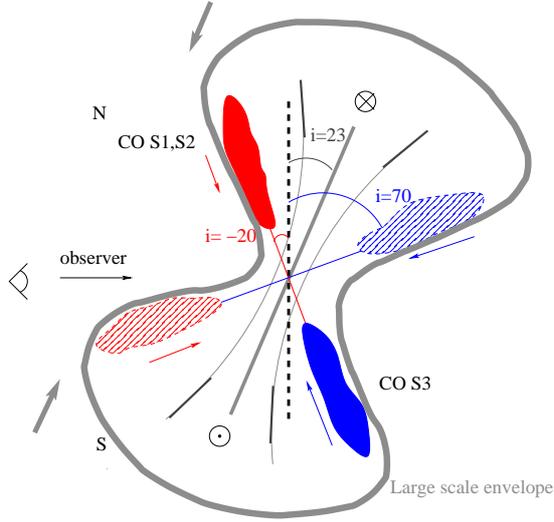}
 \caption{Illustration of the origin of the spiral emissions. The flared disk is plotted in gray segments and arcs. The solid filaments are the detected CO spirals. The hatched filaments with relatively larger inclination angles are not detected. The thick lines mark the large-scale, flattened, and tilted infalling envelope.} \label{fig:spi_incli}
\end{center}
\end{figure}

\subsection{Overall picture of the AB Aur system}\label{sec:global_picture}

Gas and dust clearly indicate the existence of an inner cavity, of  radius $\simeq 100$ AU.
As AB Aur is a strong UV emitter, a possible explanation of such a cavity could be photoevaporation. Following \citet{Alexander+etal_2006}, a transitional ``hole'' can appear during the evaporation of the disk. The disk mass in AB Aur is $\sim$ 0.01 M$_{\sun}$, which is a factor of ten more massive than in their model. In addition, the inner disk is not completely evaporated since it has been detected at NIR. The high accretion rate of AB Aur ($\sim$10$^{-7}$ $\msun$ yr$^{-1}$) furthermore suggests that the inner disk is still replenished. This inner disk can provide substantial attenuation for ionizing radiation from the central star, and photoevaporation cannot be the dominant process in clearing the disk. Besides, the asymmetry of the dust ring has no simple explanation in this scenario.
The AB Aur disk probably cannot be explained by such a model.

An alternative formation mechanism of the gap/cavity is that the disk is in tidal interaction with an embedded companion or (proto-)planet. Accounting
for the inner disk, the width of the continuum gap is $\sim$ 90 AU.
Resonant motions with a companion in the
gap can produce high-contrast inhomogeneities \citep[e.g.][]{Kuchner+etal_2003}, which would naturally explain
the observed azimuthal brightness variations in the ring. However,
the companion mass  should be high enough so that the
gap can remain mostly clear of gas or dust, as in the simulation work of \citet{Kley+etal_2001}.
Following \citet{Takeuchi+etal_1996}, the half-width of a gap, $w$, opened by a low-mass companion in the case of a circular orbit
can be estimated from:
$w = 1.3~r~A^{1/3}$, where $r$ is the semi-major axis of the companion orbit and $A$ the strength ratio of tidal to viscous effects given by 
$A=(M_p/ M_*)^2 (3\alpha(h/r)^2)^{-1}$. With standard values for the ratio of the scale height $h$ to the orbit radius $r$, $h/r \sim 0.1$
and the viscosity parameter, $\alpha, \sim 0.01$ and assuming a stellar mass, $M_*$ of  2.4 $\msun$, a gap of half-width $w$ of 45 AU located at $r \simeq 45$~AU can be opened by a body of mass $M_p \simeq 0.03 \msun$.
However, \citet{Hashimoto+etal_2011} rule out companions more massive than about five Jupiter masses from IR polarization intensity images. Based on an extreme assumption, 100\% of polarized emission for the planet, this may lead to a low value of the upper limit. Nevertheless, having one or multiple lower mass bodies in eccentric orbit or a lower viscosity (since the minimum mass scales as $\sqrt{\alpha}$) may partially alleviate these discrepancies.

Although the NIR spirals appear more consistent with a simple flared disk model, we cannot explain the CO spirals with such a model.
At NIR wavelengths, opacity prevents us from seeing throughout the disk, and we only see the front surface.
In the northern part (the farside), the front surface is projected at a lower inclination angle, while in the southern part (the nearside), it is projected at a higher inclination angle.
This results in a compression of the NIR spirals toward the south (see Fig. \ref{fig:spi_incli} for an illustration).
However, the CO opacity \textit{at the spiral velocities} is moderate, so we
could in principle see throughout the disk.
An explanation of the CO velocities in terms of different inclinations of the front and back surfaces of a flared disk would only be possible if we saw two anomalous velocity components symmetrically placed around the normal
disk velocity. The CO spirals must originate in a structure that is more complex than a flared disk.

{\it Are the CO spirals body-driven?}
It has been shown numerically that a companion in a disk is expected to drive a single-armed spiral wake \citep[e.g.][]{Ogilvie2002}. The multiple arcs/spiral arms observed in the NIR and mm appear difficult to reconcile with a single companion. Multiple bodies spread between distances ranging from 30 to 70 AU would offer a better alternative. However, the pitch angle of such spirals is typically very small.
The biggest difficulty with the body-driven spiral pattern is the apparent retrograde motion of the gas. This cannot be obtained with
purely in-plane motions. The easiest explanation for the change of the sign in velocity is an inclination angle difference. This could relate to the indications of disk warp. In the multiple-body hypothesis, resonant interaction between these bodies may pump
up orbital inclination (and eccentricity) changes, and perhaps more naturally lead to highly ($> 35^\circ$) inclined spiral patterns.
However, simulations of multiplanet + disk systems have shown that inclination and eccentricity are in general damped quite efficiently
in such cases \citep[see review by][and references therein]{Kley+Nelson_2012}. Finally, we note that density contrast in companion-driven
density waves is generally significant (10\% to 30 \% or so), while the features we have detected here in CO are much fainter.

Counter-rotating gas and spiral pattern may also trace a recent fly-by. Such an event would naturally induce a significant
warp in the disk \citep{Nixon+Pringle_2010}.
Traces of the event are, however, expected to decay on a timescale
$t_\mathrm{damp} \simeq P_\mathrm{out} ( 2/ \pi / \alpha)$ \citep{Lubow+etal_2002}, where $P_{\rm out}$ is the orbital period at the edge of the outer disk.
The $t_\mathrm{damp}$ is on the order $10^6$ yr for $\alpha = 10^{-4}$.
\citet{Pietu+etal_2005} suggest that two young stars might have had an encounter with AB Aur on such a timescale, JH433 \citep{Jones_Herbig1979}
at any time older than 35 000 years ago, and RW Aur some 500 000 years ago.
In this respect, it is interesting to note that RW Aur is a very peculiar T Tauri star binary \citep[$1.5~\msun$ for the
primary, $0.8~\msun$ for the secondary,][]{Woitas+etal_2001}, with a high accretion rate, which exhibits clear evidence of tidal interactions between the two stars, suggesting an eccentric orbit \citep{Cabrit+etal_2006}.
However, the proper motion measurement accuracy of the suggested fly-by stars precludes accurate conclusions.

 An additional important piece of information of the kinematics provided by our observations is the mis-alignment between the ``large-scale'' ($> 2000$ AU) velocity gradient
in the cloud and the AB Aur disk axis.
If interpreted by rotation, it indicates that the specific angular momentum of the surrounding envelope projects nearly perpendicular to the AB Aur disk axis. Material from this envelope should thus accrete with very different kinematics from that of the disk, perhaps explaining the apparent counter-rotating spirals.

The large-scale velocity gradient might also be interpreted by \textit{infall} motions. From the scattered light images, the southern part is closest to us, and from Fig. \ref{fig:largeco}, this gas is indeed red-shifted as expected for infall. The CO spirals could be traces of enhanced
accretion along filaments.  In the simplest model, which assumes infall from an envelope with rigid-body rotation, accreting material is expected to follow quasi-parabolic, nonintersecting orbits \citep{Cassen_Moosman1981}. In this case, our ``spirals'' may just trace some parts of such parabolic orbits, and their apparent geometry remains consistent with such an interpretation.  In the orbit plane, the pitch angle of the ``spirals'' is given by the ratio of radial to rotational velocity

\begin{equation}
\tan (\theta_{\rm pit}) =  s~V_{\rm rad} / V_{\rm rot},
\end{equation}
where $s$ is a sign that depends on conventions for velocities. This intrinsic pitch angle can be recovered from the ``spiral'' geometry by assuming some inclination angle $i$ for the orbit plane. For  $|i| < 30^\circ$, the correction for inclination will remain small.   The shape of our
``spirals'' is consistent with apparent counter-clockwise rotation (i.e. direct rotation in the left-handed RA,Dec system) and inward motions.
For $i \simeq -20^\circ$, i.e. an orbit plane inclined by about $45^\circ$ compared to the main quasi-Keplerian disk of AB Aur at $i = +23^\circ$, this direction of rotation is consistent with the overall rotation of the system.  The infall motions then contribute to red-shifted gas along
the line of sight in the northern regions (CO Spiral S1 and S2) and blue-shifted gas in the southern part (CO S3).
For S4, which lies
close to the disk's major axis, the projection of the infall motions would essentially be zero.
This behavior is entirely consistent
with the observed velocity shifts in Fig. \ref{fig:spectra}. In addition, the $y_{\rm kep}$ values of the $\dco~_{\rm spiral}$ component follow the trend of a pure Keplerian rotation 
with $M$ of 2.4 $\msun$ and $i$ of $-20\degr$ (Fig. \ref{fig:spa_v_3spec}), suggesting that the spiral gas develops from a higher inclination with respect to the disk plane. The analysis combining both rotation and radial motion suggests that the gas motion within the spirals are mainly rotation with slightly radial motion.
A schematic is given in Fig. \ref{fig:spi_incli}.

We do not see evidence of any gas coming from the opposite side of the main disk (dashed clumps in Fig. \ref{fig:spi_incli}).
This may be the result of the combinations of two effects. First, the ``spirals'' are visible because they trace density fluctuations.
Such fluctuations are inherently random, so we just observe one snapshot of events. Second, similar features located on the opposite
side would be much more inclined along the line of sight, around $70^\circ$. Thus, they would only project much closer to the center of
mass and at relatively high velocities.
As infall gives a velocity gradient along the project minor axis, while rotation leads to a velocity gradient along the major axis, combination of this gas with the main rotating disk would lead to a distortion of the iso-velocity pattern,
which could be difficult to identify if the amount of infalling gas is small as in the identified spirals (clumps in solid lines in Fig. \ref{fig:spi_incli}).

The accretion rate through the spiral can be estimated by dividing the spiral mass by the free-fall timescale at 1000 AU;
this gives about $10^{-10}$ to $10^{-8}~\msun$ yr$^{-1}$. This must be understood as a lower limit for the total envelope accretion rate, since
gas with no specific overdensity is not accounted for in this process.

Infall motions are common in Class 0 protostellar envelopes, \citep[e.g. L1527][]{Ohashi+etal_1997}, \citep[L1544][]{Tafalla+etal_1998}, as suggested by the predominance of blue-shifted asymmetries in the line profiles observed at moderate angular resolutions \citep{Mardones+etal_1997}. Nonaxisymmetric accretion has also been invoked by \citet{Tobin+etal_2012} to explain the kinematic structure of Class 0 protostellar envelopes seen at high angular resolution \citep{Tobin+etal_2011}. The AB Aur case is special, however, because the estimated age of the star is 1 to 4 Myr \citep[see discussion in][]{Pietu+etal_2005}. As the free-fall time scale at 10 000 AU is on the order of 10$^5$ years, and scales at most as $r^{3/2}$ (when the envelope mass is negligible), material still accreting at this stage should come from initial
distances around 50 000 AU, unless some mechanism has slowed down the infalling material. The change in magnetic support due to ambipolar
diffusion is a possible candidate \citep[see e.g.][and references therein]{Mouschovias_1996}.

AB Aur is for the time being the oldest object in which infall motions may have been detected. However, the apparent peculiarity of AB Aur in this respect may be a selection effect. AB Aur is a relatively massive star, and most likely the most massive of all stars with disks already studied in the Taurus clouds. Most studies of disks around young stars have instead been performed on stars isolated enough from their molecular cloud to reduce confusion and focus on the disk properties. Accordingly, very little is known about the history of accretion from the
envelopes at large ages.

\section{Summary}
We present in this paper new mm continuum and CO observations with an angular resolution as high as 0$\farcs$37 (50 AU). The key results are summarized below.

\begin{enumerate}
\item Unresolved 1.3 \,mm emission was detected toward the star. The spectral index
   between 3.6 cm and 1.3 \,mm is consistent with optically thick, free-free emission in a jet.
   However, the 1.3\,mm emission may also trace an inner dust disk, of radius smaller than  $50 $ AU,
   and (minimal)  gas+dust mass, $M_{\rm gas}, \sim$4$\times$10$^{-5}$ M$_{\sun}$.

\item A dust gap with a width of $\sim$ 90 AU was then clearly detected. It can be formed with a companion mass 0.03 M$_{\sun}$ at a radius 45 AU. Since the outer disk is highly asymmetric and more massive than the disk mass predicted by the photoevaporation model by a factor of 10, our results are more in favor of the existence of a companion.

\item At a radius of $\sim$145 AU from the star, a resolved dust ring was observed. We find that this ring is inhomogeneous in both radius (lower-middle panel of Fig. \ref{fig:continuum}) and azimuth (Fig. \ref{fig:azimuth}). The 1.3 mm flux density is 80 mJy, corresponding to $M_{\rm gas}$ of 5$\times$10$^{-3}$ $\msun$.

\item Emission of the CO \jdu~line was also observed within the dust ring extending inside the dust cavity down to a radius $\leq$ 20 AU. The axis of this CO inner disk is at a positional angle of -38$\degr$ (Table \ref{tab:cohigh}). This CO gas is likely rotating ,but at non-Keplerian velocities
   with an inclination angle of 29$\degr$, which is intermediate between the large-scale inclination of 42$\degr$ and the inclination angle of $\le$ 20$\degr$ derived from NIR interferometric measurements. This suggests that the structure is warped. The origin of this inner CO disk is probably multiple. This may be gas located in the inner part of the ring contaminated by possible streamers flowing onto the inner dust disk through the dust cavity.

\item Four CO spirals with lengths $\sim$ 500 AU were detected based on the integrated intensity and the velocity dispersion maps. The $M_{\rm gas}$ is 10$^{-7}$ $<$ $M_{\rm gas}$ $<$ 10$^{-5}$ M$_{\sun}$ and the kinetic energy, $E_{\rm kin}$, contained in the spirals  is 10$^{36}$ $<$ $E_{\rm kin}$ $<$ 10$^{38}$ erg (Sect. \ref{section:gas_mass}). In comparison to the disk mass of 0.01 M$_{\sun}$ and the possible companion, $M_{\rm gas}$ and $E_{\rm kin}$ in the spirals are very small. A small perturbation of the companion with a mass of 0.03 M$_{\sun}$ at a radius of 45 AU can release enough energy.
However, the multiple spirals are difficult to explain with a single perturbing body.
Two of the CO spirals (CO S2 and CO S3) have their infrared counterparts (S1 and S3). The base of the CO S2 is seen to be part of the dust ring (Fig. \ref{fig:over_spi}).


\item The kinematics of the excess \dco~\jdu~ emission on the spirals were also analyzed. We found that this small amount of gas is apparently counter-rotating with respect to the rotation of the ring/disk system. Such behavior is difficult to explain through planet formation scenario. A hypothetic fly-by (perhaps the nearby binary RW Aur) may lead to this unusual situation.

\item Using the large-scale velocity gradient measured with the 30-m, we suggest that the most likely explanation for the spirals and their peculiar kinematics is inhomogeneous gas accreting well above and/or below the midplane of the main disk. This explains the apparent counter-rotation by a significant change ($\sim 45^\circ$) in inclination angle. It  would indicate that weak accretion could still occur 1 to 4 Myr after the initial star formation.

\end{enumerate}

We acknowledge the anonymous referee for the thorough comments that significantly helped to make the manuscript clearer. This research was supported by NSC grants NSC99-2119-M-001-002-MY4 and NSC98-2119-M-001-024-MY4. This work was supported by ``Programme National de Physique Stellaire'' (PNPS) and ``Programme
National de Physique Chimie du Milieu Interstellaire'' (PCMI) from INSU/CNRS.

\bibliographystyle{aa}
\bibliography{abaur_sg}

\end{document}